\documentclass[journal]{IEEEtran}
\usepackage{lineno}
\usepackage{cite}
\usepackage{hyperref}
\usepackage{amsmath,amssymb,amsfonts}
\usepackage{amsmath}
\usepackage{amsthm}
\DeclareMathOperator*{\argmax}{argmax}

\usepackage{graphicx}
\usepackage{textcomp}
\usepackage{xcolor}
\usepackage{graphicx}
\usepackage{float}
\usepackage{amsmath}
\usepackage{amsfonts,amssymb}
\usepackage{mathrsfs}
\usepackage{mathtools}
\usepackage{multirow}
\usepackage{algpseudocode}
\usepackage{hyperref}
\usepackage{comment}

\usepackage[inline, shortlabels]{enumitem}
\makeatletter
\newif\if@restonecol

\usepackage{algpseudocode}
\usepackage{algorithm}

\usepackage{caption}
\usepackage{subcaption}
\usepackage{ragged2e}
\usepackage{orcidlink}
\DeclareMathOperator*{\argmin}{argmin}
\usepackage{mathtools}
\usepackage{dsfont}
\usepackage{bbm}
\usepackage{stfloats}
\usepackage{xcolor}
\usepackage{bm}

\newtheorem{corollary}{Corollary}

\theoremstyle{definition}
\newtheorem{theorem}{Theorem}

\newtheorem{lemma}{Lemma}

\newcommand{\biggg}{\bBigg@{3}}
\newcommand{\Biggg}{\bBigg@{3.5}}
\makeatother
\begin{document}

\title{Rotatable and Movable Antenna-Enabled Near-Field Integrated Sensing and Communication}
\author{Yunan~Sun, Hao~Xu,~\IEEEmembership{Graduate~Student~Member,~IEEE,} Chongjun~Ouyang, and Hongwen~Yang,~\IEEEmembership{Member,~IEEE}
%\thanks{This paper was support by 5G Evolution Wireless Air interface Intelligent R\&D and Verification Public Platform Project (Grant. 2022-229-220).
%\textit{(Corresponding author: Hao~Xu; Hongwen~Yang.)} }
\thanks{Y. Sun, H. Xu and H. Yang are with the School of Information and Communication Engineering, Beijing University of Posts and Telecommunications, Beijing, 100876, China (e-mail: \{sunyunan, Xu\_Hao, yanghong\}@bupt.edu.cn).}
\thanks{C. Ouyang is with the School of Electronic Engineering and Computer Science, Queen Mary University of London, London E1 4NS, U.K., and also with the School of Electrical and Electronic Engineering, The University of Manchester, Manchester, U.K. (e-mail: c.ouyang@qmul.ac.uk).}
%\thanks{(Corresponding author: Hongwen~Yang)}
}

%\author{Yunan~Sun
%\thanks{Y. Sun is with the School of Information and Communication Engineering, Beijing University of Posts and Telecommunications, Beijing, 100876, China (e-mail: sunyunan@bupt.edu.cn).}
%\thanks{C. Ouyang is with the School of Electrical and Electronic Engineering, University College Dublin, Dublin 4, D04 V1W8 Ireland, and also with the School of Electronic Engineering and Computer Science, Queen Mary University of London, E1 4NS London, U.K. (e-mail: chongjun.ouyang@ucd.ie).}
%\thanks{(Corresponding author: Hongwen~Yang)}
%}

\maketitle

\begin{abstract}
 The aim of this article is to investigate the performance of near-field integrated sensing and communication (ISAC) systems using rotatable movable antennas (RMAs). In the proposed RMA-enabled system, the positions and rotations of antennas at the base station (BS) are dynamically adjusted to enhance both communication and sensing capabilities. Two designs are explored: i) a sensing-centric design that minimizes the Cram$\text{\'e}$r-Rao bound (CRB) with signal-to-interference-plus-noise ratio (SINR) constraints, and ii) a communication-centric design that maximizes the sum-rate with a CRB constraint. To solve the formulated optimization problems, two alternating optimization (AO)-based algorithms are proposed capitalizing on the semidefinite relaxation (SDR) method and the particle swarm optimization (PSO) method. Numerical results demonstrate that: i) the proposed RMA-enabled system outperforms the conventional fixed-position antenna and non-rotatable movable antenna systems in both sensing-centric and communication-centric designs and RMAs' rotations show a higher performance gain in communication-centric design; ii) the proposed optimization methods achieve the Pareto boundary in both sensing-centric and communication-centric designs.
\end{abstract}


\section{Introduction}
Integrated Sensing and Communication (ISAC) is a promising technology in the evolution of the sixth generation (6G) wireless systems, which converges the traditionally separate domains of communication and sensing into a unified framework \cite{Zhang2021}. Due to its ability to efficiently allocate time, frequency, power, and hardware resources for simultaneous communication and sensing tasks \cite{Liu2022}, ISAC has been officially approved by IMT-2030 as one of six decisive usage scenarios for 6G networks and attracted increasing research attention. To further enhance the performance of sensing and communication, ISAC has been combined with multiple-antenna technologies, such as multiple-input multiple-output (MIMO) \cite{Ouyang2023,Zhou2025}, to exploit additional spatial degrees of freedom (DoFs) and achieve greater array gains for both communication and sensing tasks. These improvements contribute to higher network throughput and improved sensing resolution. In general, the performance gains offered by antenna arrays scale approximately linearly with the aperture size. This insight has inspired the development of large-aperture array architectures, including reconfigurable intelligent surfaces (RIS) \cite{Zhou2024,Luo2022}, dynamic metasurfaces \cite{Shlezinger2021}, holographic MIMO \cite{Adhikary2024}, and other related technologies, which have been widely adopted in ISAC systems.

However, the shift toward large-aperture arrays represents more than a mere quantitative increase in array size; it necessitates a paradigm shift from conventional far-field transmission to near-field transmission. Specifically, the expansion of antenna aperture enlarges the near-field region—an aspect often overlooked in traditional wireless systems. When user terminals (UTs) or sensing targets are located within the near-field region, electromagnetic wave propagation must be modeled using spherical wavefronts rather than the planar wave approximation typically used in far-field scenarios \cite{Liu2023}. This transition calls for a fundamental reexamination of existing signal processing algorithms and system designs, thereby stimulating a growing body of research on near-field ISAC \cite{Zhao2024,Zhao20242}.

\subsection{Prior Works}
Benefiting from near-field propagation, near-field ISAC can obtain accurate resolution in spatial sensing, such as joint sensing of targets' angle and distance information, while ensuring high communication performance. {\color{black}Recent studies have demonstrated the feasibility and of acquiring three-dimensional (3D) information through near-field sensing and ISAC techniques \cite{Hua2024,Hua20242}. Nevertheless, most near-field ISAC systems are based on conventional fixed-position antennas (FPAs). Such deployment limits the diversity and spatial multiplexing performance of MIMO systems because the channel variation in the spatial field is not fully utilized \cite{Zhu20240}.}

To overcome the inherent limitations of FPAs, movable antennas (MAs) and six-dimensional MAs (6DMAs) are introduced to wireless networks, which can fully exploit the spatial DoFs of antenna arrays in communication or sensing scenarios \cite{Zhu2024,Zhu20242,Shao2024,Shao20242}. MA-enabled systems can dynamically adjust the positions of antennas, thereby reshaping the channel conditions to boost communication performance, or reconfiguring the geometric properties of the antenna arrays to enhance sensing capability \cite{Wu2024}. Building upon MAs, 6DMA leverages the DoFs in both the 3D positioning and rotation of antenna planar. Due to the advantageous properties of MA and 6DMA, there has recently been extensive research on various wireless sensing and ISAC systems enabled by these technologies \cite{Wu2024,Qin2024,Xiang2024,Kuang2024,Ding2024,Shao20243}. The authors in \cite{Wu2024} investigated a MA-enabled ISAC system with RIS and maximized the minimum beampattern gain by jointly optimizing the transmit beamforming, the positions of MAs, and the phase shifts of RIS. In addition, the authors in \cite{Qin2024} studied the Cram$\text{\'e}$r-Rao bound (CRB) minimization problem for a MA-enabled ISAC system. Based on existing research on the MA-enabled far-field ISAC systems \cite{Xiang2024,Kuang2024}, the authors in \cite{Ding2024} investigated the application of MAs in a dual-functional full-duplex ISAC system with near-field channel model. Moreover, the authors in \cite{Shao20243} proposed a wireless sensing system enabled by 6DMA and minimized the CRB for estimating the directions of arrival by jointly optimizing the 6DMAs’ positions and rotations at the base station (BS), comparing it with MAs for directive and isotropic antenna radiation patterns.

However, it is worth noting that the performance of the MA planes' rotations in near-field ISAC remains unexplored, especially considering that the existing literature has employed a spherical wave model, which fails to account for the loss in channel gain resulting from the effective antenna aperture \cite{Lu2022,Dardari2020}. On the other hand, although 6DMA can achieve significant gains in both communication and sensing performance, its multiple rotatable surfaces require high hardware costs.
\subsection{Contributions}
Motivated by the above issues, we propose a rotatable MAs (RMAs) enabled near-field ISAC framework, which can be regarded as a 6DMA-enabled system with movable elements on only two planes. The main contributions of this paper are summarized as follows:
\begin{enumerate}[label=\arabic*)]
\item We propose an RMA-enabled near-field ISAC framework. In the proposed scenario, the BS is equipped with two rotatable MA planes for transmission and reception, respectively, and all the MAs can move in large-size regions, serving multiple downlink UTs as well as performing sensing on one target.
\item We investigate the near-field ISAC performance by considering the sensing-centric design and the communication-centric design. For the sensing-centric design, we formulate a CRB minimization problem with signal-to-interference-plus-noise (SINR) constraints. Regarding the communication-centric design, we formulate a sum-rate maximization problem with SINR constraints and a CRB constraint.
\item We propose two AO-based algorithms to solve the formulated problems. Specifically, we optimize the transmit beamforming and the RMAs' positions and rotations by leveraging the semidefinite relaxation (SDR) method and the particle swarm optimization (PSO) method for the sensing-centric problem. Additionally, we derive a more trackable form of the communication-centric problem by utilizing the quadratic transform and optimize the transmit beamforming and RMAs' positions and rotations by adopting similar methods for the sensing-centric problem.
\item Numerical results are present to demonstrate the convergence and effectiveness of the proposed algorithms. We also compare the performance of different setups for antennas at BS to study the individual effect and synergy of RMAs' rotations and element position movement. Moreover, we utilize the simulation results of discrete rotations to analyze the performance loss with a given number of quantization bits.
\end{enumerate}
\subsection{Organization and Notations}
The rest of this paper is organized as follows. In Section \ref{s2}, our proposed conceptual framework of the RMA-enabled near-field ISAC system model is described. In Section \ref{s3}, a CRB minimization problem for sensing-centric design is formulated, and an AO-based algorithm is proposed to solve it. In Section \ref{s4}, a sum-rate maximization problem for communication-centric design is characterized, and the corresponding AO-based algorithm is proposed. Numerical results are provided to verify the effectiveness of the proposed framework in Section \ref{s5}, and a concise conclusion is provided in Section \ref{s6}.

\textit{Notations:} Throughout this paper, scalars, vectors, and matrices are denoted by non-bold, bold lower-case, and bold upper-case letters, respectively. For the vector $\mathbf{a}$, $[\mathbf{a}]_i$, $\mathbf{a}^{\mathsf T}$, $\mathbf{a}^*$, and $\mathbf{a}^{\mathsf H}$ denote the $i$-th entry, transpose, conjugate, and conjugate transpose of $\mathbf{a}$, respectively. The notations $|{a}|$ and $\|\mathbf{b}\|$ denote the magnitude and norm of scalar $a$ and vector $\mathbf{b}$, respectively. $\mathbf{A} \succeq \mathbf{0}$ means that matrix $\mathbf{A}$ is positive semidefinite; $\mathbf{A} \succeq \mathbf{B}$ means that $\mathbf{A-B} \succeq \mathbf{0}$; $\text{rank}(\mathbf{A})$ and $\text{tr}(\mathbf{A})$ denote the rank and trace of matrix $\mathbf{A}$, respectively; $\Re\{\cdot\}$ and $\Im\{\cdot\}$ denote the real and imaginary component of a complex number, respectively.
The $N \times N$ identity matrix is denoted by $\mathbf{I}_N$. The set $\mathbb{C}^{N\times M}$ and $\mathbb{R}^{N\times M}$ stands for the space of ${N\times M}$ complex and real matrices, respectively. Finally, $\mathcal{CN}(\mu, \mathbf{X})$ is used to denote the circularly symmetric complex Gaussian distribution with mean $\mu$ and covariance matrix $\mathbf{X}$.

\section{System Model}\label{s2}
\subsection{RMA-BS Model}
\begin{figure}[!t]
\centering
\setlength{\abovecaptionskip}{-6pt}
\includegraphics[height=0.32\textwidth]{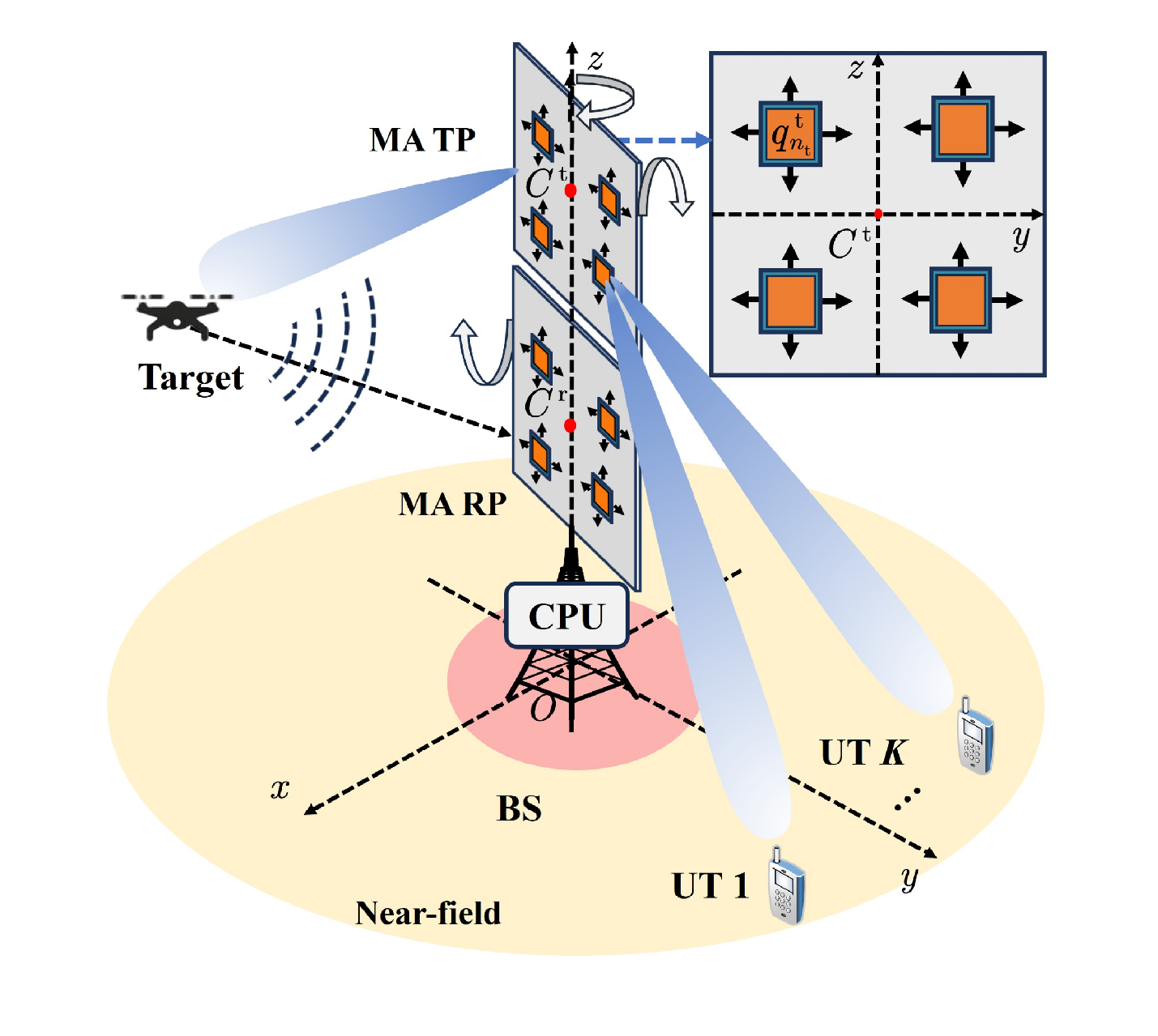}
\captionsetup{font={small}}
\caption{\justifying RMA-enabled near-field ISAC system.}
\label{Model}
\end{figure}

We consider an RMA-enabled near-field ISAC framework, where a monostatic BS with RMAs communicates with $K$ single-FPA UTs while simultaneously sensing a nearby target, as depicted in Fig. \ref{Model}. The BS is equipped with $N_{\text t}$ transmit MAs and $N_{\text r}$ receive MAs that can move in two 2D planes, namely the transmit plane (TP) and the receive plane (RP), respectively. Without loss of generality, the MAs are assumed to move within the $y$-$z$ plane of the local Cartesian coordinate system defined by the center of the corresponding plane. {\color{black}Therefore, the local positions of the $n_{\text t}$-th transmit MA and the $n_{\text r}$-th receive MA can be respectively represented as follows:
\begin{align}
%\mathbf{q}_{n_{\text t}}^{\text t}&=[0,y_{n_{\text t}}^{\text t},z_{n_{\text t}}^{\text t}]^{\mathsf T}\in\mathcal{C}^{\text t},\\
%\mathbf{q}_{n_{\text r}}^{\text r}&=[0,y_{n_{\text r}}^{\text r},z_{n_{\text r}}^{\text r}]^{\mathsf T}\in\mathcal{C}^{\text r},
\mathbf{q}_{n_{\text p}}^{\text p}&=[0,y_{n_{\text p}}^{\text p},z_{n_{\text p}}^{\text p}]^{\mathsf T}\in\mathcal{C}^{\text p},
\end{align}where ${\text p}\in\{\text{t,r}\}$; $\mathcal{C}^{\text t}$ and $\mathcal{C}^{\text r}$ denote the square moving regions of size $D\times D$ with the center points at the origin of corresponding local coordinates.}

Both the TP and the RP are connected to the CPU via rotatable rods embedded with flexible wires, and thus can execute 3D rotations. {\color{black}The rotations of TP and RP can be respectively characterized by
\begin{align}
%\bm{\varpi}^{\text t}&=[\alpha^{\text t},\beta^{\text t},\gamma^{\text t}]^{\mathsf T},\\
%\bm{\varpi}^{\text r}&=[\alpha^{\text r},\beta^{\text r},\gamma^{\text r}]^{\mathsf T},
\bm{\varpi}^{\text p}&=[\alpha^{\text p},\beta^{\text p},\gamma^{\text p}]^{\mathsf T},
\end{align}where $\alpha^{\text p}\in[0,2\pi)$, $\beta^{\text p}\in[0,2\pi)$, and $\gamma^{\text p}\in[0,2\pi)$ denote the rotation angles with respect to (w.r.t.) the $x$-axis, $y$-axis, and $z$-axis, respectively, with ${\text p}\in\{\text{t,r}\}$.}

Given $\bm{\varpi}^{\text t}$ and $\bm{\varpi}^{\text r}$, the rotation matrix can be defined as follows \cite{Shao2024}:
\begin{align}
\mathbf{F}\left(\bm{\varpi}^{\text p}\right)\triangleq\begin{bmatrix}
\cos{\alpha^{\text p}}\cos{\gamma^{\text p}}&\cos{\alpha^{\text p}}\sin{\gamma^{\text p}}&-\sin{\alpha^{\text p}}\\
F_{21}\left(\bm{\varpi}^{\text p}\right)&F_{22}\left(\bm{\varpi}^{\text p}\right)&\cos{\alpha^{\text p}}\sin{\beta^{\text p}}\\
F_{31}\left(\bm{\varpi}^{\text p}\right)&F_{32}\left(\bm{\varpi}^{\text p}\right)&\cos{\alpha^{\text p}}\cos{\beta^{\text p}}
\end{bmatrix},\nonumber
\end{align}where
\begin{align}
F_{21}\left(\bm{\varpi}^{\text p}\right)=\sin{\beta^{\text p}}\sin{\alpha^{\text p}}\cos{\gamma^{\text p}}-\cos{\beta^{\text p}}\sin{\gamma^{\text p}}, \\ F_{22}\left(\bm{\varpi}^{\text p}\right)=\sin{\beta^{\text p}}\sin{\alpha^{\text p}}\sin{\gamma^{\text p}}+\cos{\beta^{\text p}}\cos{\gamma^{\text p}}, \\ F_{31}\left(\bm{\varpi}^{\text p}\right)=\cos{\beta^{\text p}}\sin{\alpha^{\text p}}\cos{\gamma^{\text p}}+\sin{\beta^{\text p}}\sin{\gamma^{\text p}}, \\ F_{32}\left(\bm{\varpi}^{\text p}\right)=\cos{\beta^{\text p}}\sin{\alpha^{\text p}}\sin{\gamma^{\text p}}-\sin{\beta^{\text p}}\cos{\gamma^{\text p}}.
\end{align}As a result, the positions of the $n_{\text t}$-th transmit MA and the $n_{\text r}$-th receive MA in the global Cartesian coordinate system can be expressed as follows:
\begin{align}
\mathbf{t}_{n_{\text t}}&=\mathbf{F}^{\text t}\mathbf{q}_{n_{\text t}}^{\text t}+\mathbf{C}^{\text t},\\
\mathbf{r}_{n_{\text r}}&=\mathbf{F}^{\text r}\mathbf{q}_{n_{\text r}}^{\text r}+\mathbf{C}^{\text r},
\end{align}where $\mathbf{F}^{\text p}=\mathbf{F}\left(\bm{\varpi}^{\text p}\right)$; $\mathbf{C}^{\text t}$ and $\mathbf{C}^{\text r}$ denote the centers' positions of TP and RP, respectively.

Additionally, we consider two practical constraints on RMA.
\subsubsection{Rotation Constraints to Avoid Signal Reflection}As per \cite{Shao2024}, the TP and RP must meet some rotation constraints to avoid mutual signal reflections between them, which can be expressed as follows:
\begin{align}
&\mathbf{u}^{\text t}\left(\mathbf{r}_{n_{\text r}}-\mathbf{C}^{\text t}\right)\leq 0,\forall n_{\text r}\in[1,N_{\text r}],\label{rc1}\\
&\mathbf{u}^{\text r}\left(\mathbf{t}_{n_{\text t}}-\mathbf{C}^{\text r}\right)\leq 0,\forall n_{\text t}\in[1,N_{\text t}],\label{rc2}
\end{align}with $\mathbf{u}^{\text p}=\mathbf{F}^{\text p}\left[1,0,0\right]^{\mathsf T}$ denoting the outward normal vector of the TP or RP. This ensures that all the antennas are positioned within the halfspace that consists of vectors forming obtuse angles with the normal vector of another plane.
\subsubsection{Minimum-Distance Constraint}To avoid potential electrical coupling between MAs, a minimum inter-MA distance $d_{\min}$ is required among MAs \cite{Zhu2023}, i.e.,
\begin{align}
&\|\mathbf{t}_{n_{\text t}}-\mathbf{t}_{n_{\text t}^\prime}\|\geq d_{\min},n_{\text t}\neq n_{\text t}^\prime,n_{\text t},n_{\text t}^\prime\in[1,N_{\text t}],\label{dc1}\\
&\|\mathbf{t}_{n_{\text r}}-\mathbf{t}_{n_{\text r}^\prime}\|\geq d_{\min},n_{\text r}\neq n_{\text r}^\prime,n_{\text r},n_{\text r}^\prime\in[1,N_{\text r}].\label{dc2}
\end{align}

\subsection{Channel Model}
In this paper, we investigate downlink multiuser transmission with a quasi-static channel model and assume that the UTs and the sensing target are located in the near-field region of the BS. The positions of the UT/target are given by $\mathbf{p}_k=[d_k\sin{\theta_k}\cos{\phi_k},d_k\sin{\theta_k}\sin{\phi_k},d_k\cos{\theta_k}]^{\mathsf T}$, with $d_k$ denoting the distance between the UT/target and the global origin, $\theta_k\in[0,\pi]$ denoting the elevation angle, and $\phi_k\in[-\frac{\pi}{2},\frac{\pi}{2}]$ denoting the azimuth angle, where $k\in[0,K]$ with $k=0$ representing the sensing target and $k\in[1,K]$ representing the $k$-th UT. Then the distance from the $k$-th UT/target to the $n_{\text t}$ transmit antenna and the $n_{\text r}$ receive antenna can be respectively calculated by $\Delta^{\text t}_{k,n_{\text t}}=\|\mathbf{p}_k-\mathbf{t}_{n_{\text t}}\|$ and $\Delta^{\text r}_{k,n_{\text r}}=\|\mathbf{p}_k-\mathbf{r}_{n_{\text r}}\|$.

We assume that both the users and the target are located within the near-field region of the RMA, owing to its large aperture size. As discussed earlier, this necessitates the use of a spherical wavefront-based model to accurately characterize the wireless channel. Specifically, in the context of spherical-wave propagation, the signals received at different antennas arrive from distinct angles and distances. Consequently, the effective aperture of each antenna element varies depending on its spatial position relative to the source, which significantly impacts the array response and beamforming behavior. The effective antenna aperture is determined by the product of the maximum value of the effective area and the projection of the array normal to the direction of the signal, with the resultant loss in channel power gain referred to as effective aperture loss \cite{Lu2022}. Therefore, under free-space line-of-sight propagation, the channel power gain between the $n_{\text p}$-th antenna and the $k$-th UT/target is given by \cite{Zhao20242}
\begin{align}
|{h}^{\text p}_{n_{\text p},k}|^2=\int_{\mathcal{S}_{n_\text p}}\mathcal{L}^{\text p}(\mathbf{p}_k,\mathbf{s})\mathcal{G}^{\text p}(\mathbf{p}_k,\mathbf{s})d\mathbf{s},\label{h2}
\end{align}where $\mathcal{S}_{n_\text p}$ denotes the region of $n_{\text p}$ antenna with size of $\sqrt{A}\times\sqrt{A}$, and
\begin{align}
\mathcal{L}^{\text p}(\mathbf{p}_k,\mathbf{s})&=\frac{1}{4\pi\|\mathbf{p}_k-\mathbf{s}\|^2},\\
\mathcal{G}^{\text p}(\mathbf{p}_k,\mathbf{s})&=\frac{\left(\mathbf{p}_k-\mathbf{s}\right)^{\mathsf T}\mathbf{u}^{\text p}\delta\left(\mathbf{p}_k,\mathbf{s},\mathbf{u}^{\text p}\right)}{\|\mathbf{p}_k-\mathbf{s}\|},
\end{align} denote the free-space path loss and the effective aperture loss, respectively, with $\delta\left(\mathbf{p}_k,\mathbf{s},\mathbf{u}^{\text p}\right)$ representing the sign function w.r.t. $\left(\mathbf{p}_k-\mathbf{s}\right)^{\mathsf T}\mathbf{u}^{\text p}$ for guaranteeing non-negativity of the projection. Given that the size of each antenna is significantly smaller than the propagation distance, i.e., $d_k\gg\sqrt{A}, \forall k\in[0,K]$, the variation of the channel among different points $\mathbf{s}\in\mathcal{S}_{n_\text p}$ is negligible. Therefore, \eqref{h2} can be approximated as
\begin{align}
|{h}^{\text t}_{n_{\text t},k}|&=\sqrt{A\mathcal{L}^{\text t}(\mathbf{p}_k,\mathbf{t}_{n_{\text t}})\mathcal{G}^{\text t}(\mathbf{p}_k,\mathbf{t}_{n_{\text t}})},\\
|{h}^{\text r}_{n_{\text r},k}|&=\sqrt{A\mathcal{L}^{\text r}(\mathbf{p}_k,\mathbf{r}_{n_{\text r}})\mathcal{G}^{\text r}(\mathbf{p}_k,\mathbf{r}_{n_{\text r}})},
\end{align} for ${\text p}={\text t}$ and ${\text r}$, respectively. As a result, the near-field channel between the $k$-th UT/target and the $n_{\text p}$ antenna can be derived as
\begin{align}
{h}^{\text p}_{n_{\text p},k}=|{h}^{\text p}_{n_{\text p},k}|{\text e}^{-{\text j}\frac{2\pi}{\lambda}\Delta^{\text p}_{k,n_{\text p}}},\label{nfc_h}
\end{align}where $\lambda=\frac{c}{f_{\text c}}$ is the signal wavelength with $f_{\text c}$ denoting the carrier frequency and $c$ denoting the speed of propagation. In contrast to conventional far-field channel, the phase in \eqref{nfc_h} is related to the distance between the transmit and receive antennas, rather than merely to the angle between them, and it is a nonlinear function of antennas' positions, rather than a linear one.

\subsection{Signal Model}
We consider a coherent time block of $T$, during which the parameters for communication and sensing remain constant. The transmitted signal from the BS at time $t\in[1,T]$ can be modeled as
\begin{align}
\mathbf{x}(t)=\mathbf{W}\mathbf{s}[t]+\mathbf{x}_0(t)=\sum_{k=1}^{K}\mathbf{w}_ks_k(t)+\mathbf{x}_0(t),
\end{align}where $\mathbf{W}=[\mathbf{w}_1,\dots,\mathbf{w}_K]\in\mathbb{C}^{N_{\text t}\times K}$ denotes the digital beamformer for conveying the information symbol $\mathbf{s}(t)=\left[s_1(t),\dots,s_K(t)\right]^{\mathsf{T}}\in\mathbb{C}^{K\times 1}$ to the UTs and $\mathbf{x}_0(t)\in\mathbb{C}^{N_{\text t}\times 1}$ denotes the dedicate sensing signal to achieve the full DoFs for target sensing \cite{Liu2020}. The multiple beam transmission is also exploited by the dedicated sensing signal, whose covariance matrix $\mathbf{R}_0=\mathbb{E}\left[\mathbf{x}_0(t)\mathbf{x}^{\mathsf H}_0(t)\right]$ is of a general rank. Moreover, the information symbols are modeled as independent Gaussian random variables with zero mean and unit power while the dedicated sensing signal is generated by pseudo-random coding, so that $\mathbb{E}\left[\mathbf{s}(t)\mathbf{s}^{\mathsf H}(t)\right]=\mathbf{I}_K$ and $\mathbb{E}\left[\mathbf{s}(t)\mathbf{x}^{\mathsf H}_0(t)\right]=\mathbf{0}_{K\times N_{\text t}}$. Thus the covariance matrix of the transmit signal $\mathbf{x}(t)$ is given by
\begin{align}
\mathbf{R}_x=\mathbb{E}\left[\mathbf{x}(t)\mathbf{x}^{\mathsf H}(t)\right]=\mathbf{W}\mathbf{W}^{\mathsf H}+\mathbf{R}_0.
\end{align}
As per \cite{Wang2023}, $\mathbf{R}_x$ can be approximately calculated by
\begin{align}
\mathbf{R}_x\approx\frac{1}{T}\mathbf{X}\mathbf{X}^{\mathsf H},\label{approx}
\end{align}where $\mathbf{X}=[\mathbf{x}(1),\dots,\mathbf{x}(T)]$. This approximation is accurate when $T$ is large enough. In this paper, we assume that \eqref{approx} holds accurate equal.

Based on the aforementioned channel and transmitted signal model, the received signal at the $k$-th UT for communication is given by
\begin{align}
y_k(t)=\sum_{i=1}^{K}\mathbf{h}_k^{\mathsf T}\mathbf{w}_is_i(t)+\mathbf{h}_k^{\mathsf T}\mathbf{x}_0(t)+n_k(t),\label{yk}
\end{align}where ${\mathbf{h}_k}=[{h}^{\text t}_{1,k};\dots,{h}^{\text t}_{N_{\text t},k}]^{\mathsf T}$ and $n_k(t)\sim\mathcal{CN}(0,\sigma_k^2)$ denotes the additive Gaussian white noise (AWGN) with $\sigma_k^2$ representing the noise power. And the received echo signal at the BS for target sensing is given by
\begin{align}
\mathbf{y}_0(t)=\mathbf{G}\mathbf{x}(t)+\mathbf{n}_0(t),
\end{align}where $\mathbf{G}=\eta\mathbf{g}_0\mathbf{h}_0^{\mathsf T}$ with $\eta$ denoting the complex channel gain of the sensing target, ${\mathbf{g}_0}=[{h}^{\text r}_{1,0};\dots,{h}^{\text r}_{N_{\text t},0}]^{\mathsf T}$, and $\mathbf{n}_0(t)\sim\mathcal{CN}(\mathbf{0}_{N_{\text r}},\sigma_0^2\mathbf{I}_{N_{\text r}})$ denotes the AWGN.

\subsection{Sensing Performance Metric: CRB}
In the near-field region, the spherical wave propagation enables joint estimation of distance and angles of the target, which can be observed in \eqref{nfc_h}. In this paper, we aim to estimate the target's positional information, i.e., the distance $d_0$, elevation angle $\theta_0$, and azimuth angle $\phi_0$ and resort to the CRB as the performance metric since CRB delineates a lower bound on the variance of unbiased estimators \cite{Kay1993}. Denote $\bm{\zeta}=[d_0,\theta_0,\phi_0,\eta^{\text r},\eta^{\text i}]$ as the vector of unknown parameters, where $\eta^{\text r}=\Re\{\eta\}$ and $\eta^{\text i}=\Im\{\eta\}$. Then the Fisher information matrix (FIM) for estimating $\bm{\zeta}$ can be partitioned as
\begin{align}
\mathbf{J}_{\bm{\zeta}}=\begin{bmatrix}
\mathbf{J}_{11}&\mathbf{J}_{12}\\
\mathbf{J}_{12}^{\mathsf T}&\mathbf{J}_{22}
\end{bmatrix},
\end{align}where $\mathbf{J}_{11}\in\mathbb{C}^{3\times 3}$ denotes the partition merely about $d_0,\theta_0$, and $\phi_0$; $\mathbf{J}_{22}\in\mathbb{C}^{2\times 2}$ denotes the partition merely about $\eta$; and $\mathbf{J}_{12}\in\mathbb{C}^{3\times 2}$ denotes the partition about the mutual information between the position information and $\eta$. The detailed expressions of $\mathbf{J}_{11},\mathbf{J}_{22}$, and $\mathbf{J}_{12}$ are derived in Appendix \ref{App1}. According to Woodbury matrix identity \cite{Woodbury1967}, the CRB matrix for estimating $d_0,\theta_0$, and $\phi_0$, which is the $3\times3$ submatrix in the upper-left corner of $\mathbf{J}_{\bm{\zeta}}^{-1}$, is given by
\begin{align}
{\text {CRB}}\left(\mathbf{R}_x,\mathbf{G},\sigma_0^2\right)=\left(\mathbf{J}_{11}-\mathbf{J}_{12}\mathbf{J}_{22}^{-1}\mathbf{J}_{12}^{\mathsf T}\right)^{-1}\label{CRB}.
\end{align}

\subsection{Communication Performance Metric: Communication Rate}
In this paper, we assume that the perfect channel state information (CSI) of communication channels is perfectly known at the BS via a proper channel estimation mechanism \cite{Xiao2024,Zhou20252}. From \eqref{yk}, the received SINR at the $k$-th UT can be expressed as follows:
\begin{align}
\Gamma_k=\frac{|\mathbf{h}_k^{\mathsf T}\mathbf{w}_k|^2}{\sum_{i\neq k}|\mathbf{h}_k^{\mathsf T}\mathbf{w}_i|^2+\mathbf{h}_k^{\mathsf T}\mathbf{R}_0\mathbf{h}_k^{\mathsf *}+\sigma_k^2}.
\end{align}Thereby, the downlink communication rate of $k$-th user is given by
\begin{align}
R_k\left(\mathbf{W},\mathbf{R_0},\mathbf{H}\right)=\log_2\left(1+\Gamma_k\right),
\end{align}where $\mathbf{H}=[\mathbf{h}_1,\dots,\mathbf{h}_K]$. Then the downlink sum-rate is given by
\begin{align}
R\left(\mathbf{W},\mathbf{R_0},\mathbf{H}\right)=\sum_{k=1}^{K}R_k\left(\mathbf{W},\mathbf{R_0},\mathbf{H}\right).
\end{align}
\section{Sensing-Centric Design}\label{s3}
\subsection{Problem Formulation}
In this section, we consider a sensing-centric design by optimizing the sensing performance while guaranteeing a certain communication performance, which can be modeled as a CRB-minimization problem with communication rate constraints. Due to the monotonic increment of logarithmic function, the communication rate constraint for each user can be considered as a SINR constraint. Consequently, the corresponding SINR-constrained CRB-minimization problem is formulated as
\begin{subequations}\label{P1}
\begin{align}
{\mathcal{P}}_{\text{1}}:&~\min_{\mathbf{W},\mathbf{R}_0,\mathbf{Q}^{\text t},\mathbf{Q}^{\text r},\bm{\varpi}^{\text t},\bm{\varpi}^{\text r}}~{\text{tr}}\left({\text {CRB}}\left(\mathbf{R}_x,\mathbf{G},\sigma_0^2\right)\right)\label{P1obj}\\
{\text{s.t.}}&~{\text{tr}}(\mathbf{R}_x)\leq P_{\max},\label{pc}\\
&~\mathbf{R}_x-\sum_{k=1}^{K}\mathbf{w}_k\mathbf{w}_k^{\mathsf H}\succeq \mathbf{0},\label{Rc}\\
&~\Gamma_k\geq \Gamma_{\min},\forall k\in[1,K],\label{SINRc}\\
&~\mathbf{q}_{n_{\text t}}^{\text t}\in\mathcal{C}^{\text t},\forall{n_{\text t}}\in[1,N_{\text t}],\label{qtc}\\
&~\mathbf{q}_{n_{\text r}}^{\text r}\in\mathcal{C}^{\text r},\forall{n_{\text r}}\in[1,N_{\text r}],\label{qrc}\\
&~\eqref{rc1},\eqref{rc2},\eqref{dc1},\eqref{dc2},\nonumber
\end{align}
\end{subequations}where $\mathbf{Q}^{\text p}=[\mathbf{q}_1,\dots,\mathbf{q}_{N_{\text p}}]$; $P_{\max}\geq 0$ is the total transmit power budget; and $\Gamma_{\min}\geq 0$ represents the minimum SINR for the $k$-th user. Since $\mathcal{P}_1$ is non-convex and the variables are highly coupled, we adopt an AO-based framework to optimize ISAC signal's covariance matrix $\mathbf{R}_x$ as well as the RMAs' positions and rotations in the following.

\subsection{Optimization of Transmit Beamforming}
In this subsection, we optimize the covariance matrix of the transmit ISAC signal at BS, i.e., $\mathbf{R}_x$, with the positions and rotation of RMAs fixed. Due to the complexity of the objective function in CRB form, we first transform the original problem into the following equivalent but more tractable form according to the Schur complement condition \cite{Wang2023}:
\begin{subequations}\label{P2}
\begin{align}
{\mathcal{P}}_{\text{2}}:&~\min_{\mathbf{W},\mathbf{R}_0,\mathbf{U}}~{\text{tr}}\left(\mathbf{U}^{-1}\right)\\
{\text{s.t.}}&~\begin{bmatrix}\mathbf{J}_{11}-\mathbf{U}&\mathbf{J}_{12}\\
\mathbf{J}_{12}^{\mathsf T}&\mathbf{J}_{22}\end{bmatrix}\succeq\mathbf{0},\label{Uc}\\
&~\eqref{pc}-\eqref{SINRc},\nonumber
\end{align}
\end{subequations}where $\mathbf{U}\in\mathbb{C}^{3\times3}$ is an auxiliary matrix. By this way, we transform the non-convex objective function into the convex constraint \eqref{Uc}. Moreover, we adopt the SDR method to deal with the non-convex constraints \eqref{Rc} and \eqref{SINRc}. Specifically, we define $\bm{\Omega}_k=\mathbf{w}_k\mathbf{w}_k^{\mathsf H}$, which is a semidefinite matrix and has a rank of $1$. Then \eqref{Rc} can be written in the following convex form:
\begin{align}
\mathbf{R}_x\succeq\sum_{k=1}^{K}\bm{\Omega}_k.\label{Rc2}
\end{align}Similarly, \eqref{SINRc} can be transformed into a convex form as:
\begin{align}
\frac{1+\Gamma_{\min}}{\Gamma_{\min}}\mathbf{h}_k^{\mathsf T}\bm{\Omega}_k\mathbf{h}_k^{*}\geq\mathbf{h}_k^{\mathsf T}\mathbf{R}_x\mathbf{h}_k^{*}+\sigma_k^2.\label{SINRc2}
\end{align}Through omitting the rank-one constraint of $\bm{\Omega}_k$, we can formulate the SDR problem as:
\begin{subequations}\label{P3}
\begin{align}
{\mathcal{P}}_{\text{3}}:&~\min_{\bm{\Omega}_k\succeq\mathbf{0},\mathbf{R}_x\succeq\mathbf{0},\mathbf{U}\succeq\mathbf{0}}~{\text{tr}}\left(\mathbf{U}^{-1}\right)\\
{\text{s.t.}}&~\eqref{pc},\eqref{Uc},\eqref{Rc2},\eqref{SINRc2}.\nonumber
\end{align}
\end{subequations}Note that $\mathcal{P}_3$ is a convex semidefinite programming (SDP) problem, the global optimum of which can be obtained by the existing convex optimization solvers such as CVX and MOSEK. Although we omit the rank-one constraint of $\bm{\Omega}_k$, the rank-one global optimal $\mathbf{w}_k$ and corresponding $\mathbf{R}_x$ of $\mathcal{P}_2$ can always be constructed from an arbitrary global optimal solution of $\mathcal{P}_3$.
\newtheorem{thm}{Proposition}
\begin{thm}\label{theorem0}
Given an arbitrary global optimum $\tilde{\mathbf{R}}_x$, $\tilde{\bm{\Omega}}_1,\dots,\tilde{\bm{\Omega}}_K$ of $\mathcal{P}_3$, the following solution is a global optimum of $\mathcal{P}_2$:
\begin{align}
&\mathbf{w}_k^\circ=\left(\mathbf{h}_k^{\mathsf T}\tilde{\bm{\Omega}}_k\mathbf{h}_k^{*}\right)^{-\frac{1}{2}}\tilde{\bm{\Omega}}_k\mathbf{h}_k^{*},\label{upw}\\
&\mathbf{R}_x^\circ=\tilde{\mathbf{R}}_x.\label{upR}
\end{align}
\end{thm}
\begin{IEEEproof}
Please refer to Appendix \ref{App2}.
\end{IEEEproof}
Then optimal $\mathbf{R}_0$ can be obtained by $\mathbf{R}_0^\circ=\mathbf{R}_x^\circ-\sum_{k=1}^{K}\mathbf{w}_k^\circ(\mathbf{w}_k^\circ)^{\mathsf H}$.

\subsection{Optimization of RMAs' Positions and Rotations}
In this subsection, we optimize the RMAs' positions and rotations, i.e., $\mathbf{Q}^{\text t},\mathbf{Q}^{\text r},\bm{\varpi}^{\text t}$, and $\bm{\varpi}^{\text r}$, with $\mathbf{R}_x$ fixed. Since the objective function is highly non-convex and the solution space is large in general, it is difficult to obtain a local optimal solution. To efficiently address this challenge, PSO is introduced as an effective approach \cite{Robinson2004,Zhu2019}.
{\color{black}In the PSO-based approach, we first initialize $M$ particles as follows:
\begin{align}
\bm{\xi}_m^{(0)}=&[{y^{\text t}}^{(0)}_{m,1},{z^{\text t}}^{(0)}_{m,1},\dots,{{y}^{\text t}}^{(0)}_{m,N_{\text t}},{{z}^{\text t}}^{(0)}_{m,N_{\text t}},{\alpha^{\text t}}^{(0)}_m,{\beta^{\text t}}^{(0)}_m,{\gamma^{\text t}}^{(0)}_m,\nonumber\\
&{y^{\text r}}^{(0)}_{m,1},{z^{\text r}}^{(0)}_{m,1},\dots,{y^{\text r}}^{(0)}_{m,N_{\text r}},{z^{\text r}}^{(0)}_{m,N_{\text r}},{\alpha^{\text r}}^{(0)}_m,{\beta^{\text r}}^{(0)}_m,{\gamma^{\text r}}^{(0)}_m]^{\mathsf T},\label{xiup}
\end{align}where ${y^{\text p}}^{(0)}_{1,n_{\text p}}=y^{\text p}_{n_{\text p}}, {z^{\text p}}^{(0)}_{1,n_{\text p}}=z^{\text p}_{n_{\text p}}$, ${\alpha^{\text p}}^{(0)}_1=\alpha^{\text p}, {\beta^{\text p}}^{(0)}_1=\beta^{\text p}, {\gamma^{\text p}}^{(0)}_1=\gamma^{\text p}$, for $n_{\text p}\in[1,N_{\text p}]$; ${y^{\text p}}^{(0)}_{m,n_{\text p}},{z^{\text p}}^{(0)}_{m,n_{\text p}}\sim\mathcal{U}(-\frac{D}{2},\frac{D}{2})$, ${\alpha^{\text p}}^{(0)}_m,{\beta^{\text p}}^{(0)}_m,{\gamma^{\text p}}^{(0)}_m\sim\mathcal{U}(0,2\pi)$ for $m\in[2,M]$, $n_{\text p}\in[1,N_{\text p}]$, with ${\text p}\in\{\text{t,r}\}$.} Then we randomly initialize each particle's velocity vector $\mathbf{v}^{(0)}_m\in\mathbb{R}^{(2N_{\text r}+2N_{\text t}+6)\times1}$ and define $\bm{\Xi}^{(0)}=\{\bm{\xi}^{(0)}_1,\dots, \bm{\xi}^{(0)}_M\}$, $\mathbf{V}^{(0)}=\{\mathbf{v}^{(0)}_1,\dots,\mathbf{v}^{(0)}_M\}$. Subsequently, we set the initial local best position of M particles as $\bm{\xi}_{m,{\text{lbest}}}=\bm{\xi}_m^{(0)}$ and select the particle with the best fitness as the global particle $\bm{\xi}_{\text{gbest}}$ of the swarm. The fitness of the particles is defined in the subsequent text.

Next, we update each particle according to its individual experience and the swarm experience, which can be described by the current local best particle $\bm{\xi}_{m,{\text{lbest}}}$ and the current global best particle $\bm{\xi}_{\text{gbest}}$. For the $\tau$-th iteration, the $m$-th particle is updated as follows:
\begin{align}
\mathbf{v}^{(\tau+1)}_m=&\omega\mathbf{v}^{(\tau)}_m+a_1b_1\left(\bm{\xi}_{m,{\text{lbest}}}-\bm{\xi}_m^{(\tau)}\right)\nonumber\\
&+a_2b_2\left(\bm{\xi}_{{\text{gbest}}}-\bm{\xi}_m^{(\tau)}\right),\label{vs}\\
\bm{\xi}_m^{(\tau+1)}=&\mathcal{B}\left(\bm{\xi}_m^{(\tau)}+\mathbf{v}^{(\tau+1)}_m\right),\label{xis}
\end{align}where $a_1$ and $a_2$ are the individual and global learning factors, which represent the step size of each particle moving toward the best position; $b_1,b_2\sim\mathcal{U}(0,1)$ are two random parameters for enhancing the randomness of the search in order to escape from local optimum; $\omega$ is the parameter for maintaining the inertia of each particle's movement. Specifically, we update $\omega$ as follows:
\begin{align}
\omega=\omega_{\max}-\frac{(\omega_{\max}-\omega_{\min})\tau}{\tau_{\max}},\label{omega}
\end{align}where $\omega_{\max}$ and $\omega_{\min}$ are the maximum and minimum values of $\omega$, respectively, with $\tau_{\max}$ denoting the maximum iteration number of PSO.
Due to the fact that the particle may move out of the feasible region, we employ a projection function to process the updated particles in \eqref{xis} as follows:
\begin{align}
[\mathcal{B}(\bm{\xi})]_i=\begin{cases}
- \frac{D}{2}, & \text{if } [\bm{\xi}]_i < -\frac{D}{2}, i\in\mathcal{Q}\\
\frac{D}{2}, & \text{if } [\bm{\xi}]_i > \frac{D}{2}, i\in\mathcal{Q}\\
[\bm{\xi}]_i, & \text{otherwise},
\end{cases}
\end{align}where $\mathcal{Q}=[1,2N_{\text t}]\cup[2N_{\text t}+4,2N_{\text t}+2N_{\text r}+3]$ represents the set of indices corresponding to the antenna positions within the particles. By utilizing $\mathcal{B}(\bm{\xi})$, constraints \eqref{qtc} and \eqref{qrc} are guaranteed to be satisfied. However, \eqref{rc1},\eqref{rc2},\eqref{dc1},\eqref{dc2}, and \eqref{SINRc} remain to be taken into consideration. To address this, we define a fitness function by introducing three penalty terms into the objective function of the original problem as follows:
\begin{align}
\mathcal{F}(\bm{\xi}_m^{(\tau)})=&{\text{tr}}({\text{CRB}}(\bm{\xi}_m^{(\tau)}))+\mu_1|\mathcal{S}(\bm{\xi}_m^{(\tau)})|\\
&+\mu_2|\mathcal{J}(\bm{\xi}_m^{(\tau)})|+\mu_3\mathcal{H}(\bm{\xi}_m^{(\tau)}),\label{Fit}
\end{align}where ${\text{CRB}}(\bm{\xi}_m^{(\tau)})$ denotes the CRB calculated by setting RMAs' positions and rotations according to $\bm{\xi}_m^{(\tau)}$; $\mu_1,\mu_2$, and $\mu_3$ are the positive penalty parameters which are large enough; $\mathcal{S}(\bm\xi)$ is a set composed of the position pairs of MAs that violate the minimum-distance constraint. Specifically, $\mathcal{S}(\bm\xi)$ is defined as follows:
\begin{align}
\resizebox{0.98\linewidth}{!}{$
\mathcal{S}(\bm\xi)=\{([\bm\xi]_n,[\bm\xi]_{n^\prime})|\|[\bm\xi]_n-[\bm\xi]_{n^\prime}\|<d_{\min},n\neq n^\prime,n,n^\prime\in\mathcal{Q}\}$}.
\end{align}Similarly, $\mathcal{J}(\bm{\xi}_m^{(\tau)})$ is a set corresponding to the rotation constraints \eqref{rc1} and \eqref{rc2}, which is defined as follows:
\begin{align}
&\mathcal{J}(\bm{\xi}_m^{(\tau)})=\left\{{n_{\text r}}|\mathbf{u}^{\text t}(\bm\xi^{(\tau)})\left(\mathbf{r}_{n_{\text r}}(\bm\xi^{(\tau)})-\mathbf{C}^{\text t}\right)>0,n_{\text r}\in[1,N_{\text r}]\right\}\nonumber\\
&\cup\left\{{n_{\text t}}|\mathbf{u}^{\text r}(\bm\xi^{(\tau)})\left(\mathbf{t}_{n_{\text t}}(\bm\xi^{(\tau)})-\mathbf{C}^{\text r}\right)>0,n_{\text t}\in[1,N_{\text t}]\right\},
\end{align}where ${\mathbf{u}^{\text p}(\bm\xi^{(\tau)})}$, ${\mathbf{t}_{n_{\text t}}(\bm\xi^{(\tau)})}$, and $\mathbf{r}_{n_{\text r}}(\bm\xi^{(\tau)})$ represent the normal vector of the TP or RP, the $n_{\text t}$-th transmit MA's position, and the $n_{\text r}$-th receive MA's position calculated based on $\bm\xi^{(\tau)}$.
Hence $|\mathcal{S}(\bm\xi)|$ and $|\mathcal{J}(\bm{\xi}_m^{(\tau)})|$ are two adaptive penalty terms \cite{Das2011}. And $\mathcal{H}(\bm{\xi}_m^{(\tau)})$ is a brick wall penalty term  corresponding to the SINR constraint \eqref{SINRc}, which is defined as follows:
\begin{align}
\mathcal{H}(\bm{\xi}_m^{(\tau)})=\sum_{k=1}^{K}\left(\max\left\{0,\Gamma_{\min}-\Gamma_k(\bm{\xi}_m^{(\tau)})\right\}\right)^2,
\end{align}where $\Gamma_k(\bm{\xi}_m^{(\tau)})$ represent the $k$-th UT's SINR calculated based on $\bm\xi^{(\tau)}$. Since the values of $\mu_1,\mu_2$, and $\mu_3$ are large enough, the penalty terms drive the particles to move to the positions where \eqref{rc1},\eqref{rc2},\eqref{dc1},\eqref{dc2}, and \eqref{SINRc} are satisfied. Thus, $|\mathcal{S}(\bm{\xi}_{\text{gbest}})|$, $|\mathcal{J}(\bm{\xi}_{\text{gbest}})|$, and $\mathcal{H}(\bm{\xi}_{\text{gbest}})$ will approach zero during the iterations so that the constraints are resultantly satisfied. By continuously selecting particles corresponding to smaller values of the fitness function, the local and global optimal particles are improved until convergence. Then a suboptimal solution for the positions and rotations of RMAs is obtained.

The detailed AO-based overall algorithm for solving $\mathcal{P}_1$ is summarized in Algorithm \ref{Algorithm1}. In line 1, $\mathbf{W}$ and $\mathbf{R}_0$ are initialized by considering the constraint \eqref{pc} while $\mathbf{Q}^{\text t},\mathbf{Q}^{\text r},\bm{\varpi}^{\text t}$, and $\bm{\varpi}^{\text r}$ are randomly initialized by considering the constraint \eqref{rc1}, \eqref{rc2}, \eqref{dc1}, \eqref{dc2}, \eqref{qtc}, and \eqref{qrc}. Then the transmit beamforming is optimized by the SDR method in line 4. Subsequently, the positions and rotations of RMAs are optimized by the PSO method in lines 6-22.
{\begin{algorithm}[t]
    \caption{AO-based algorithm for solving $\mathcal{P}_1$}
    \label{Algorithm1}\textcolor{black}{
    \begin{algorithmic}[1]
    \State Initialize $i=0$,$\mathbf{W},\mathbf{R}_0,\mathbf{Q}^{\text t},\mathbf{Q}^{\text r},\bm{\varpi}^{\text t}$, and $\bm{\varpi}^{\text r}$.
    \Repeat
    %\For{$i=1$ to $I_{\text{AO}}$}
        \State Calculate $\mathbf{H}$ and $\mathbf{G}$ with current $\mathbf{Q}^{\text t},\mathbf{Q}^{\text r},\bm{\varpi}^{\text t}$, and $\bm{\varpi}^{\text r}$.
        \State Solve $\mathcal{P}_3$ and update $\mathbf{W}$ and $\mathbf{R}_0$ by \eqref{upw} and \eqref{upR}
        %\State Calculate $f_0={\text{tr}}\left({\text {CRB}}\left(\mathbf{R}_x,\mathbf{G},\sigma_0^2\right)\right)$ by \eqref{CRB}.
        \State Initialize the M particles with positions $\bm{\Xi}^{(0)}$ and velocity $\mathbf{V}^{(0)}$.
        \State Set the local best positions $\bm{\xi}_{m,{\text{lbest}}}=\bm{\xi}_m^{(0)}$ for $m\in[1,M]$ and the global best positions $\bm{\xi}_{{\text{gbest}}}=\argmin_{\bm{\xi}_m^{(0)}}\{\mathcal{F}(\bm{\xi}_1^{(0)}),\dots,\mathcal{F}(\bm{\xi}_M^{(0)})\}$
        \For{$\tau=1$ to $\tau_{\max}$}
            \State Update the inertia parameter $\omega$ according to \eqref{omega}.
            \For{$m=1$ to M}
                \State Update the velocity and position of the $m$-th particle according to \eqref{vs} and \eqref{xis}, respectively.
                \State Calculate the fitness value of the $m$-th particle, i.e., $\mathcal{F}(\bm{\xi}_m^{(\tau)})$, according to \eqref{Fit}.
                \If{$\mathcal{F}(\bm{\xi}_m^{(\tau)})<\mathcal{F}(\bm{\xi}_{m,\text{lbest}})$}
                    \State Update $\bm{\xi}_{m,\text{lbest}}=\bm{\xi}_m^{(\tau)}$.
                \EndIf
                \If{$\mathcal{F}(\bm{\xi}_m^{(\tau)})<\mathcal{F}(\bm{\xi}_{\text{gbest}})$}
                    \State Update $\bm{\xi}_{\text{gbest}}=\bm{\xi}_m^{(\tau)}$.
                \EndIf
            \EndFor
        \EndFor
        %\If{$\mathcal{F}(\bm{\xi}_{\text{gbest}})<f_0$}
        \State Update $\mathbf{Q}^{\text t},\mathbf{Q}^{\text r},\bm{\varpi}^{\text t}$, and $\bm{\varpi}^{\text r}$ according to $\bm{\xi}_{\text{gbest}}$.
        %\EndIf
        \State Set $i=i+1$.
    \Until{Convergence or $i>I_{\max}$}
    %\EndFor
    \end{algorithmic}}
\end{algorithm}}
\subsection{Convergence and Complexity Analysis}
In Algorithm \ref{Algorithm1}, the transmit beamforming, the positions and rotations of RMAs are alternatively optimized. Note that the value of the objective function is non-increasing over the iterations. Since the trace of CRB is lower-bounded with the given transmit power, the convergence of Algorithm \ref{Algorithm1} is ensured.

{\color{black}Then we analyze the computational complexity of the proposed algorithm. For each iteration, the complexity for updating $\mathbf{W}$ and $\mathbf{R}_0$ via the interior method is in order of $\mathcal{O}(K^{6.5}N_{\text t}^{6.5}{\text{log}}\epsilon^{-1})$ with the given solution accuracy $\epsilon$ \cite{Liu2020}. And the complexity for updating $\mathbf{Q}^{\text t},\mathbf{Q}^{\text r},\bm{\varpi}^{\text t}$, and $\bm{\varpi}^{\text r}$ via the PSO method is in order of $\mathcal{O}\left(\tau_{\max}M(2N_{\text t}+2N_{\text r}+6+\text{log}M)\right)$. Therefore, the total computational complexity of Algorithm \ref{Algorithm1} is \resizebox{0.98\linewidth}{!}{$\mathcal{O}\left(I_{\text{AO}}(K^{6.5}N^{6.5}_{\text t}{\text{log}}\epsilon^{-1}+\tau_{\max}M(2N_{\text t}+2N_{\text r}+6+\text{log}M))\right)$}, with $I_{\text{AO}}$ denoting the resultant iteration number of the AO-based algorithm.}

\section{Communication-Centric Design}\label{s4}
\subsection{Problem Formulation}
In this section, we consider a communication-centric design by optimizing the communication performance while ensuring a certain sensing performance, which can be modeled as a sum-rate-maximization problem with CRB constraint. Specifically, the corresponding CRB-constraint sum-rate-maximization problem is formulated as
\begin{subequations}\label{P4}
\begin{align}
{\mathcal{P}}_{\text{4}}:&~\max_{\mathbf{W},\mathbf{R}_0,\mathbf{Q}^{\text t},\mathbf{Q}^{\text r},\bm{\varpi}^{\text t},\bm{\varpi}^{\text r}}~R\left(\mathbf{W},\mathbf{R}_0,\mathbf{H}\right)\label{P4obj}\\
{\text{s.t.}}&~{\text{tr}}\left({\text {CRB}}\left(\mathbf{R}_x,\mathbf{G},\sigma_0^2\right)\right)\leq C_{\max},\label{cc}\\
&~\eqref{rc1}-\eqref{dc2},~\eqref{pc}-\eqref{qrc},\nonumber%\eqref{rc1},\eqref{rc2},\eqref{dc1},\eqref{dc2},\eqref{pc},\eqref{Rc},\eqref{SINRc},\eqref{qtc},\eqref{qrc},\nonumber
\end{align}
\end{subequations}where $C_{\max}$ denotes the maximum tolerable value of the trace of the CRB for the sensing target's position information. It can be seen that ${\mathcal{P}}_{\text{4}}$ is also a non-convex problem with highly coupled variables. To address this challenge, we first transform it into a more tractable form and then utilize the AO-based framework to optimize the variables.
\subsection{Problem Reformulation}
Apart from the CRB constraint \eqref{cc}, $\mathcal{P}_4$ is a classic downlink sum-rate maximization problem, which can be equivalently transformed into a more tractable form by introducing the auxiliary variables $\rho_k$ for $k\in[1,K]$ as follows: \cite{Shen2018}
\begin{align}
{\mathcal{P}}_{\text{5}}:&~\max_{\mathbf{W},\mathbf{R}_0,\mathbf{Q}^{\text t},\mathbf{Q}^{\text r},\bm{\varpi}^{\text t},\bm{\varpi}^{\text r},\bm{\rho}}~f_1(\mathbf{W},\mathbf{R}_0,\mathbf{H},\bm{\rho})\label{P5obj}\\
{\text{s.t.}}&~\eqref{rc1}-\eqref{dc2},~\eqref{pc}-\eqref{qrc},~\eqref{cc},\nonumber
\end{align}where $\bm{\rho}=[\rho_1,\dots,\rho_K]^{\mathsf T}$ and
$f_1(\mathbf{W},\mathbf{R}_0,\mathbf{H},\bm{\rho})=\sum_{k}^{K}\left[\log_2(1+\rho_k)-\rho_k+\frac{(1+\rho_k)|\mathbf{h}_k^{\mathsf T}\mathbf{w}_k|^2}{\sum_{i=1}|\mathbf{h}_k^{\mathsf T}\mathbf{w}_i|^2+\mathbf{h}_k^{\mathsf T}\mathbf{R}_0\mathbf{h}_k^{\mathsf *}+\sigma_k^2}\right]$. It is obvious that ${\mathcal{P}}_{\text{5}}$ is a convex problem w.r.t. ${\rho_k}$. Hence, by checking the first-order optimality condition, the optimal ${\rho_k}$ can be obtained as follows:
\begin{align}
\rho_k^\circ=\Gamma_k. \label{rho}
\end{align}Consequently, the original problem is rewritten into the form of sum-of-ratio with $\bm\rho$ fixed.
%\begin{align}
%{\mathcal{P}}_{\text{6}}:&~\max_{\mathbf{W},\mathbf{R}_0,\mathbf{Q}^{\text t},\mathbf{Q}^{\text r},\bm{\varpi}^{\text t},\bm{\varpi}^{\text r}}~\sum_{k=1}^{K}\frac{(1+\rho_k)|\mathbf{h}_k^{\mathsf T}\mathbf{w}_k|^2}{\sum_{i=1}|\mathbf{h}_k^{\mathsf T}\mathbf{w}_i|^2+\mathbf{h}_k^{\mathsf T}\mathbf{R}_0\mathbf{h}_k^{\mathsf *}+\sigma_k^2}\label{P6obj}\\
%{\text{s.t.}}&~\eqref{rc1}-\eqref{dc2},~\eqref{pc}-\eqref{qrc},~\eqref{cc}.\nonumber
%\end{align}

Furthermore, we rewrite the current sum-of-ratio maximization problem by employing the quadratic transform proposed in \cite{Shen20182} as follows:
\begin{align}
{\mathcal{P}}_{\text{6}}:&~\max_{\mathbf{W},\mathbf{R}_0,\mathbf{Q}^{\text t},\mathbf{Q}^{\text r},\bm{\varpi}^{\text t},\bm{\varpi}^{\text r},\bm{\nu}}~f_2(\mathbf{W},\mathbf{R}_0,\mathbf{H},\bm{\rho},\bm{\nu})\label{P6obj}\\
{\text{s.t.}}&~\eqref{rc1}-\eqref{dc2},~\eqref{pc}-\eqref{qrc},~\eqref{cc},\nonumber
\end{align}where $\bm{\nu}=[\nu_1,\dots,\nu_K]^{\mathsf T}$ denotes the auxiliary variables and $f_2(\mathbf{W},\mathbf{R}_0,\mathbf{H},\bm{\rho},\bm{\nu})=\sum_{k=1}^{K}[2(1+\rho_k)\nu_k\sqrt{|\mathbf{h}_k^{\mathsf T}\mathbf{w}_k|^2}-(1+\rho_k)\nu_k^2(\mathbf{h}_k^{\mathsf T}\mathbf{R}_x\mathbf{h}_k^{\mathsf *}+\sigma_k^2)]$. Similarly, the optimal $\nu_k$ can be obtained by checking the first-order optimality condition as
\begin{align}
\nu_k^\circ=\frac{\sqrt{|\mathbf{h}_k^{\mathsf T}\mathbf{w}_k|^2}}{\mathbf{h}_k^{\mathsf T}\mathbf{R}_x\mathbf{h}_k^{\mathsf *}+\sigma_k^2}.\label{nu}
\end{align}Thus, the original problem is equivalently reformulated as:
\begin{align}
{\mathcal{P}}_{\text{7}}:&~\max_{\mathbf{W},\mathbf{R}_0,\mathbf{Q}^{\text t},\mathbf{Q}^{\text r},\bm{\varpi}^{\text t},\bm{\varpi}^{\text r}}~\sum_{k=1}^{K}\left[2(1+\rho_k)\nu_k\sqrt{\mathbf{h}_k^{\mathsf T}\mathbf{w}_k\mathbf{w}_k^{\mathsf H}\mathbf{h}_k^*}\right.\nonumber\\
&\left.\vphantom{2(1+\rho_k)\nu_k\sqrt{\mathbf{h}_k^{\mathsf T}\mathbf{w}_k\mathbf{w}_k^{\mathsf H}\mathbf{h}_k^*}}\quad\quad\quad\quad\quad\quad\quad\quad\quad-(1+\rho_k)\nu_k^2(\mathbf{h}_k^{\mathsf T}\mathbf{R}_x\mathbf{h}_k^{\mathsf *}+\sigma_k^2)\right]\label{P7obj}\\
{\text{s.t.}}&~\eqref{rc1}-\eqref{dc2},~\eqref{pc}-\eqref{qrc},~\eqref{cc}.\nonumber
\end{align}
\subsection{Optimization of Transmit Beamforming}
In this subsection, we optimize $\mathbf{R}_x$ with $\mathbf{Q}^{\text t},\mathbf{Q}^{\text r},\bm{\varpi}^{\text t}$, and $\bm{\varpi}^{\text r}$ fixed. Similar to the sensing-centric scenario, we propose to adopt the SDR method to optimize the transmit beamforming. However, \eqref{cc} is a complex non-convex constraint, which hinders us from directly using the SDR method to transform the original problem into a convex form. Fortunately, we can rewrite \eqref{cc} into a simpler generalized inequality form via the Schur complement condition as we do in the second subsection of Section \ref{s3}. To this end, we derive the following Proposition.
\begin{thm}\label{theorem1}
The inequality ${\text{tr}}\left({\text {CRB}}\left(\mathbf{R}_x,\mathbf{G},\sigma_0^2\right)\right)\leq C_{\max}$ has a sufficient condition as the following generalized inequality:
\begin{align}
\begin{bmatrix}\mathbf{J}_{11}-\frac{3}{C_{\max}}\mathbf{I}_3&\mathbf{J}_{12}\\
\mathbf{J}_{12}^{\mathsf T}&\mathbf{J}_{22}\end{bmatrix}\succeq\mathbf{0}.\label{Cc}
\end{align}
\end{thm}
\begin{IEEEproof}
By the Schur complement condition, we can derive the following generalized inequality from \eqref{Cc}:
\begin{align}
\mathbf{J}_{11}-\mathbf{J}_{12}\mathbf{J}_{22}^{-1}\mathbf{J}_{12}^{\mathsf T}\succeq\frac{3}{C_{\max}}\mathbf{I}_3,
\end{align}with $\mathbf{J}_{22}\succeq\mathbf{0}$.
According to \cite{Boyd2004}, the function ${\text{tr}}(\mathbf{A}^{-1})$ is matrix decreasing on the positive semidefinite matrix space. Therefore, we can derive the following inequality:
\begin{align}
{\text{tr}}\left((\mathbf{J}_{11}-\mathbf{J}_{12}\mathbf{J}_{22}^{-1}\mathbf{J}_{12}^{\mathsf T})^{-1}\right)\leq{\text{tr}}(\frac{C_{\max}}{3}\mathbf{I}_3)=C_{\max},
\end{align}which is equivalent to \eqref{cc}.
\end{IEEEproof}{\color{black}Note that \eqref{cc} has a necessary and sufficient condition in the form of \eqref{Uc}, which does not guarantee a sufficiently small CRB for each element of the target's position information. In this paper, we adopt the stricter condition \eqref{Cc} to ensure the sensing accuracy and reliability.}

Based on Proposition \ref{theorem1}, we formulate the SDR problem similarly to the second subsection of Section \ref{s3} as:
\begin{align}
{\mathcal{P}}_{\text{8}}:&~\max_{\bm{\Omega}_k\succeq\mathbf{0},\mathbf{R}_x\succeq\mathbf{0}}~\sum_{k=1}^{K}2(1+\rho_k)\nu_k\sqrt{\mathbf{h}_k^{\mathsf T}\bm{\Omega}_k\mathbf{h}_k^*}\nonumber\\
&\quad\quad\quad\quad\quad\quad\quad-(1+\rho_k)\nu_k^2(\mathbf{h}_k^{\mathsf T}\mathbf{R}_x\mathbf{h}_k^{\mathsf *}+\sigma_k^2)\label{P8obj}\\
%{\text{s.t.}}&~\begin{bmatrix}\mathbf{J}_{11}(\mathbf{R}_x)-\frac{3}{C_{\max}}\mathbf{I}_3&\mathbf{J}_{12}(\mathbf{R}_x)\\
%\mathbf{J}_{12}(\mathbf{R}_x)^{\mathsf T}&\mathbf{J}_{22}(\mathbf{R}_x)\end{bmatrix}\succeq\mathbf{0},\\
&~\eqref{pc},\eqref{Rc2},\eqref{SINRc2},\eqref{Cc}.\nonumber
\end{align}Also, $\mathcal{P}_8$ is a convex SDP problem, which can be solved by the existing convex optimization solvers. The corresponding optimal $\mathbf{w}_k^\circ$ and $\mathbf{R}_0^\circ$ can be obtained by \eqref{upw} and \eqref{upR}, respectively.
\subsection{Optimization of RMAs’ Positions and Rotations}
In this subsection, we optimize $\mathbf{Q}^{\text t},\mathbf{Q}^{\text r},\bm{\varpi}^{\text t}$, and $\bm{\varpi}^{\text r}$, with $\mathbf{R}_x$ fixed. Since the constraint \eqref{cc} is highly non-convex and the solution space is generally large, it is difficult to obtain a local optimal solution. Similar to the third subsection of Section \ref{s3}, we propose to leverage the PSO-based approach.

We generate $M$ particles and their velocity vectors according to \eqref{xiup} and initialize the local best positions $\bm{\xi}_{m,{\text{lbest}}}$ as well as the global best position $\bm{\xi}_{{\text{gbest}}}$. Additionally, we calculate the inertia of each particle’s movement for each iteration by \eqref{omega} and update each particle's velocity vector by \eqref{vs}. Due to the fact that the particle may move out of the feasible region, we still need to employ the projection function to make sure that \eqref{qtc} and \eqref{qrc} are satisfied. Then each particle is updated by \eqref{xis}.

Particularly, we define a fitness function by introducing four penalty terms into the objective function of the original problem as follows:
\begin{align}
\mathcal{E}(\bm{\xi}_m^{(\tau)})=&f_3(\bm{\xi}_m^{(\tau)})-\mu_1|\mathcal{S}(\bm{\xi}_m^{(\tau)})|-\mu_2|\mathcal{J}(\bm{\xi}_m^{(\tau)})|\nonumber\\
&-\mu_3\mathcal{H}(\bm{\xi}_m^{(\tau)})-\mu_4\mathcal{T}(\bm{\xi}_m^{(\tau)}),\label{Fit2}
\end{align}where $f_3(\bm{\xi}_m^{(\tau)})=\sum_{k=1}^{K}2(1+\rho_k)\nu_k\sqrt{|\mathbf{h}_k(\bm{\xi}_m^{(\tau)})^{\mathsf T}\mathbf{w}_k|^2}-(1+\rho_k)\nu_k^2(\mathbf{h}_k(\bm{\xi}_m^{(\tau)})^{\mathsf T}\mathbf{R}_x\mathbf{h}_k(\bm{\xi}_m^{(\tau)})^{\mathsf *}+\sigma_k^2)$ with $\mathbf{h}_k(\bm{\xi}_m^{(\tau)})$ denoting the channel from BS to the $k$-th UT $\mathbf{h}_k$ calculated according to the current particle position $\bm{\xi}_m^{(\tau)}$; $\mathcal{T}(\bm{\xi}_m^{(\tau)})$ is a brick wall penalty term corresponding to the CRB constraint \eqref{cc}, which is defined as
\begin{align}
\mathcal{T}(\bm{\xi}_m^{(\tau)})=\left(\max\left\{0,{\text{tr}}\left({\text{CRB}}(\bm{\xi}_m^{(\tau)})\right)-C_{\max}\right\}\right)^2.
\end{align}Moreover, since the value of $\mathcal{T}(\bm{\xi}_m^{(\tau)})$ is often numerically much smaller than $f_3(\bm{\xi}_m^{(\tau)})$, we set an adaptive penalty parameter $\mu_4$ as
\begin{align}
\mu_4=\frac{\mu_0}{{\text{tr}}\left({\text {CRB}}\left(\mathbf{R}_x,\mathbf{G},\sigma_0^2\right)\right)^2},\label{mu4}
\end{align}where $\mu_0$ is a positive standard penalty parameter comparable to other penalty parameters. It means that we dynamically adjust $\mu_4$ based on the value of the CRB's trace computed by the updated transmit beamforming. Then all the penalty parameters are large enough. Hence, $|\mathcal{S}(\bm{\xi}_{\text{gbest}})|$, $|\mathcal{J}(\bm{\xi}_{\text{gbest}})|$, $\mathcal{H}(\bm{\xi}_{\text{gbest}})$, and $\mathcal{T}(\bm{\xi}_{\text{gbest}})$ will approach zero during the iterations so that \eqref{rc1},\eqref{rc2},\eqref{dc1},\eqref{dc2}, \eqref{SINRc}, and \eqref{cc} are guaranteed to be satisfied. By continuously selecting particles corresponding to larger values of the fitness function, the local and global optimal particles are improved until convergence. Then a suboptimal solution for the positions and rotations of RMAs is obtained.

The detailed AO-based overall algorithm for solving $\mathcal{P}_4$ is summarized in Algorithm \ref{Algorithm2}. In line 1, $\mathbf{W}$ and $\mathbf{R}_0$ are initialized by considering the constraint \eqref{pc} while $\mathbf{Q}^{\text t},\mathbf{Q}^{\text r},\bm{\varpi}^{\text t}$, and $\bm{\varpi}^{\text r}$ are randomly initialized by considering the constraint \eqref{rc1}, \eqref{rc2}, \eqref{dc1}, \eqref{dc2}, \eqref{qtc}, and \eqref{qrc}. The auxiliary variables are updated in line 4. Then the transmit beamforming is optimized by the SDR method in line 5. Subsequently, the positions and rotations of RMAs are optimized by the PSO method in lines 7-25.
\setlength{\textfloatsep}{0pt}
\setlength{\intextsep}{0pt}
\setlength{\floatsep}{0pt}
\begin{algorithm}[t]
    \caption{AO-based algorithm for solving $\mathcal{P}_4$}
    \label{Algorithm2}
    \textcolor{black}{\begin{algorithmic}[1]
    \State Initialize $i=0$,$\mathbf{W},\mathbf{R}_0,\mathbf{Q}^{\text t},\mathbf{Q}^{\text r},\bm{\varpi}^{\text t}$, and $\bm{\varpi}^{\text r}$.
    \Repeat
    %\For{$i=1$ to $I_{\text{AO}}$}
        \State Calculate $\mathbf{H}$ and $\mathbf{G}$ with current $\mathbf{Q}^{\text t},\mathbf{Q}^{\text r},\bm{\varpi}^{\text t}$, and $\bm{\varpi}^{\text r}$.
        \State Update $\bm{\rho}$ and $\bm{\nu}$ by \eqref{rho} and \eqref{nu}, respectively.
        \State Solve $\mathcal{P}_8$ and update $\mathbf{W}$ and $\mathbf{R}_0$ by \eqref{upw} and \eqref{upR}.
        %\State Calculate $e_0=f_2(\mathbf{W},\mathbf{R}_0,\mathbf{H},\bm{\rho},\bm{\nu})$.
        \State Initialize the M particles with positions $\bm{\Xi}^{(0)}$ and velocity $\mathbf{V}^{(0)}$.
        \State Update $\mu_4$ by \eqref{mu4}.
        \State Set the local best positions $\bm{\xi}_{m,{\text{lbest}}}=\bm{\xi}_m^{(0)}$ for $m\in[1,M]$ and the global best positions $\bm{\xi}_{{\text{gbest}}}=\argmax_{\bm{\xi}_m^{(0)}}\{\mathcal{E}(\bm{\xi}_1^{(0)}),\dots,\mathcal{E}(\bm{\xi}_M^{(0)})\}$
        \For{$\tau=1$ to $\tau_{\max}$}
            \State Update the inertia parameter $\omega$ according to \eqref{omega}.
            \For{$m=1$ to M}
                \State Update the velocity and position of the $m$-th particle according to \eqref{vs} and \eqref{xis}, respectively.
                \State Calculate the fitness value of the $m$-th particle, i.e., $\mathcal{E}(\bm{\xi}_m^{(\tau)})$, according to \eqref{Fit2}.
                \If{$\mathcal{E}(\bm{\xi}_m^{(\tau)})>\mathcal{E}(\bm{\xi}_{m,\text{lbest}})$}
                    \State Update $\bm{\xi}_{m,\text{lbest}}=\bm{\xi}_m^{(\tau)}$.
                \EndIf
                \If{$\mathcal{E}(\bm{\xi}_m^{(\tau)})>\mathcal{E}(\bm{\xi}_{\text{gbest}})$}
                    \State Update $\bm{\xi}_{\text{gbest}}=\bm{\xi}_m^{(\tau)}$.
                \EndIf
            \EndFor
        \EndFor
        %\If{$\mathcal{E}(\bm{\xi}_{\text{gbest}})>e_0$}
        \State Update $\mathbf{Q}^{\text t},\mathbf{Q}^{\text r},\bm{\varpi}^{\text t}$, and $\bm{\varpi}^{\text r}$ according to $\bm{\xi}_{\text{gbest}}$.
        %\EndIf
        \State Set $i=i+1$.
    \Until{Convergence or $i>I_{\max}$}
    %\EndFor
    \end{algorithmic}}
\end{algorithm}
\subsection{Convergence and Complexity Analysis}
In Algorithm \ref{Algorithm2}, the transmit beamforming, the positions and rotations of RMAs are alternatively optimized. The value of sum-rate is non-decreasing over the iterations. Due to the equivalent objective property of the two transforms in problem reformulation, the introduction of auxiliary variables does not change the non-decreasing trend of the objective functions. Since the sum-rate is upper-bounded with a given transmit power budget $P_{\max}$, Algorithm \ref{Algorithm2} is ensured to converge.

Next, we analyze the computational complexity of the proposed algorithm. For each iteration, the complexity of updating $\bm\rho$ and $\bm\nu$ are $\mathcal{O}(K)$. And as calculated in the last subsection of Section \ref{s3}, the complexity for updating $\mathbf{W}$ and $\mathbf{R}_0$ via the interior method is in order of $\mathcal{O}(K^{6.5}N_{\text t}^{6.5}{\text{log}}\epsilon^{-1})$ with the given solution accuracy $\epsilon$. The complexity for updating $\mathbf{Q}^{\text t},\mathbf{Q}^{\text r},\bm{\varpi}^{\text t}$, and $\bm{\varpi}^{\text r}$ via the PSO method is in order of $\mathcal{O}\left(\tau_{\max}M(2N_{\text t}+2N_{\text r}+6+\text{log}M)\right)$. Therefore, the total computational complexity of Algorithm \ref{Algorithm2} is \resizebox{0.98\linewidth}{!}{$\mathcal{O}\left(I_{\text{AO}}(2K+K^{6.5}N_{\text t}^{6.5}{\text{log}}\epsilon^{-1}+\tau_{\max}M(2N_{\text t}+2N_{\text r}+6+\text{log}M))\right)$}, with $I_{\text{AO}}$ denoting the resultant iteration number of the AO-based algorithm.

\section{Numerical Results}\label{s5}
In this section, we evaluate the two designs in Section \ref{s3} and Section \ref{s4} and demonstrate the effectiveness of our proposed algorithms for sensing and communication performance optimization.
\subsection{Simulation Setup}
We consider a global Cartesian coordinate system with the origin located at the BS. The center of TP and RP are set as $\mathcal{C}^{\text t}=[0,0,5+D/2]^{\mathsf T}$ and $\mathcal{C}^{\text r}=[0,0,5-D/2]^{\mathsf T}$, respectively. The MAs at the BS are confined to a square moving region, which is modeled as a rectangle within the planar local Cartesian coordinate system relative to the center of TP/RP, with local coordinates ranging from $[-D/2,D/2]\times[-D/2,D/2]$. The communication users are assumed to be randomly distributed in the $x-y$ plane, specifically $d_k\sim\mathcal{U}(15,25)$, $\theta_k=\frac{\pi}{2}$, and $\phi_k\sim\mathcal{U}(-\frac{\pi}{2},\frac{\pi}{2})$ for $k=[1,K]$. The sensing target's position is set as $d_0=10$, $\theta_0=\frac{\pi}{3}$, and $\phi_0=\frac{\pi}{4}$. The simulation parameters are detailed in Table \ref{table1}, unless stated otherwise. It should be noted that due to the large aperture size, the target and the users are located within the Rayleigh distance.
{\begin{table}[t]
\caption{Simulation Parameters}
%\captionsetup{justification=raggedright,singlelinecheck=false}
\small
\centering
\begin{tabular}{|c|p{4cm}|c|}
\hline
\textbf{Parameter} & $~~~~~~~~~~~$\textbf{Description} & \textbf{Value} \\
\hline
$K$ & Number of users & 4 \\
\hline
$N_{\text t}$ & Number of transmit MAs & 9 \\
\hline
$N_{\text r}$ & Number of receive MAs & 9 \\
\hline
$f_{\text c}$ & Carrier frequency & 24GHz \\
\hline
$d_{\min}$ & Minimum inter-MA distance & $\lambda/2$ \\
\hline
$D$ & Length of MAs' moving region  & $80\lambda$ \\
\hline
\textcolor{black}{$A$} & \textcolor{black}{Size of MA elements}  &\textcolor{black} {$\frac{\lambda^2}{4\pi}$\cite{Zhao2024} }\\
\hline
$P_{\max}$ & Transmit power budget & 40dB \\
\hline
\textcolor{black}{$\sigma_k^2,\sigma_0^2$} &\textcolor{black}{Noise power of users\& receiver} & \textcolor{black}{-110dBm} \\
\hline
\multirow{2}{*}{$I_{\max}$} & Maximum iteration number in Algorithm \ref{Algorithm1} and Algorithm \ref{Algorithm2} & \multirow{2}{*}{20} \\
\hline
\multirow{2}{*}{$M$} &Number of particles in Algorithm \ref{Algorithm1} and Algorithm \ref{Algorithm2}&\multirow{2}{*}{200} \\
\hline
\multirow{2}{*}{$\tau_{\max}$} &Maximum PSO iteration in Algorithm \ref{Algorithm1} and Algorithm \ref{Algorithm2}&\multirow{2}{*}{100} \\
\hline
\multirow{2}{*}{\textcolor{black}{$a_1$}} &\textcolor{black}{Individual learning factor in Algorithm \ref{Algorithm1} and Algorithm \ref{Algorithm2}}&\multirow{2}{*}{\textcolor{black}{1.4\cite{Ding2025}}} \\
\hline
\multirow{2}{*}{\textcolor{black}{$a_2$}} &\textcolor{black}{Global learning factor in Algorithm \ref{Algorithm1} and Algorithm \ref{Algorithm2}}&\multirow{2}{*}{\textcolor{black}{1.4\cite{Ding2025}}} \\
\hline
\multirow{2}{*}{\textcolor{black}{$\omega_{\min}$}} &\textcolor{black}{Minimum inertial weight in Algorithm \ref{Algorithm1} and Algorithm \ref{Algorithm2}}&\multirow{2}{*}{\textcolor{black}{0.4\cite{Ding2025}}} \\
\hline
\multirow{2}{*}{\textcolor{black}{$\omega_{\max}$}} &\textcolor{black}{Maximum inertial weight in Algorithm \ref{Algorithm1} and Algorithm \ref{Algorithm2}}&\multirow{2}{*}{\textcolor{black}{0.9\cite{Ding2025}}} \\
\hline
\multirow{2}{*}{$\mu_{0}$} &Standard penalty parameter in Algorithm \ref{Algorithm2}&\multirow{2}{*}{1000} \\
\hline
\multirow{2}{*}{$\mu_{1}$} &Penalty parameter in Algorithm \ref{Algorithm1} and Algorithm \ref{Algorithm2}&\multirow{2}{*}{1000} \\
\hline
\multirow{2}{*}{$\mu_{2}$} &Penalty parameter in Algorithm \ref{Algorithm1} and Algorithm \ref{Algorithm2}&\multirow{2}{*}{1000} \\
\hline
\multirow{2}{*}{$\mu_{3}$} &Penalty parameter in Algorithm \ref{Algorithm1} and Algorithm \ref{Algorithm2}&\multirow{2}{*}{1000} \\
\hline
\end{tabular}
\label{table1}
\end{table}}
\begin{figure}[!t]
\centering
\begin{subfigure}[b]{\linewidth}
    \includegraphics[width=\linewidth]{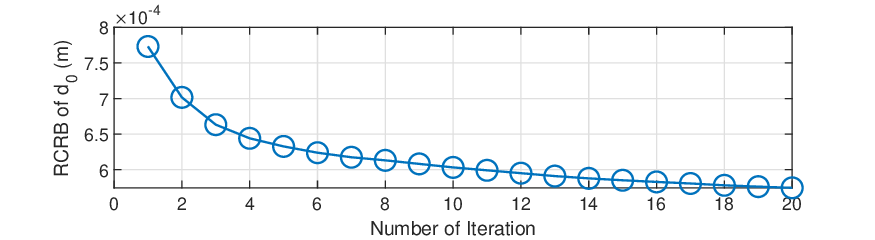}
    \caption{RCRB of $d_0$.}
    \label{cd}
\end{subfigure}
\begin{subfigure}[b]{\linewidth}
    \includegraphics[width=\linewidth]{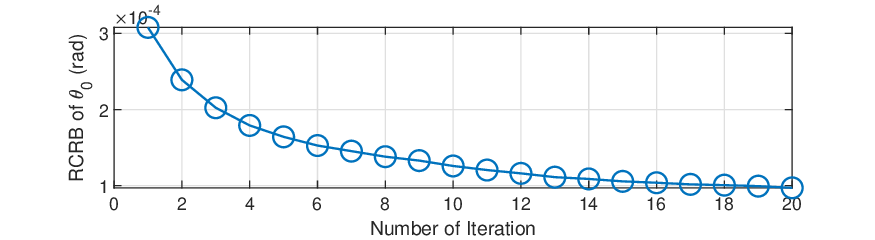}
    \caption{RCRB of $\theta_0$.}
    \label{ctheta}
\end{subfigure}
\begin{subfigure}[b]{\linewidth}
    \includegraphics[width=\linewidth]{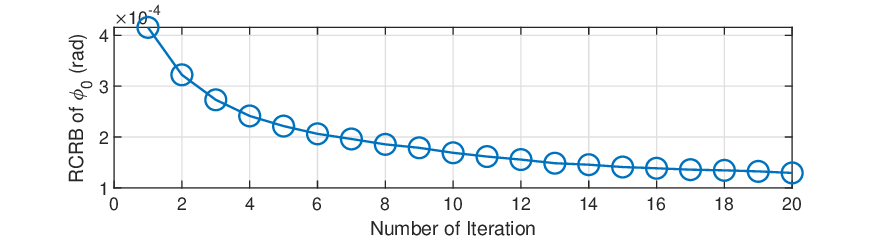}
    \caption{RCRB of $\phi_0$.}
    \label{cphi}
\end{subfigure}
\begin{subfigure}[b]{\linewidth}
    \includegraphics[width=\linewidth]{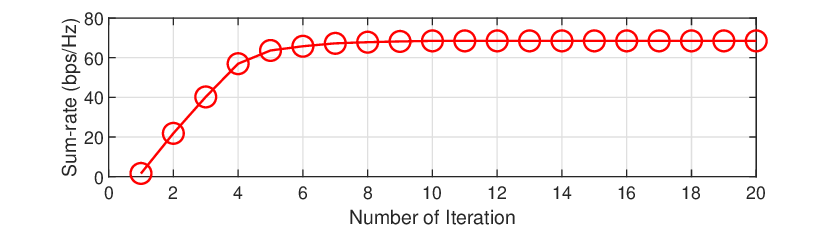}
    \caption{Sum-rate.}
    \label{csr}
\end{subfigure}
\caption{\justifying Performance versus the iteration number.}
\label{c}
\end{figure}
\subsection{Convergence Performance}
Fig. \ref{c} presents the convergence performance of the proposed AO-based algorithms for the sensing-centric and communication-centric designs in the RMA-aided near-field ISAC system. In Fig. \ref{c}, we set the minimum SINR for each UT as $\Gamma_{\min}=6$dB and the maximum trace of CRB for the sensing target as $C_{\max}=1\times10^{-6}$. As depicted in Fig. \ref{cd}-\ref{cphi}, Algorithm \ref{Algorithm1} converges within about $20$ iterations for solving the sensing-centric problem. The root CRB (RCRB) of the distance information $d_0$ decreases from $8.2\times10^{-4}$ to $5.7\times10^{-4}$; the RCRB of the elevation angle information $\theta_0$ from $3.1\times10^{-4}$ to $9.0\times10^{-5}$; the RCRB of the azimuth angle information $\theta_0$ from $4.2\times10^{-4}$ to $1.2\times10^{-4}$, which confirms the effectiveness of our proposed algorithm for obtaining feasible solutions and improving sensing performance. Note that in the first iteration, Algorithm \ref{Algorithm1} optimizes the transmit beamforming as well as the positions and rotations of RMAs, satisfying the minimum SINR constraints. In Fig. \ref{csr}, Algorithm \ref{Algorithm2} converges within about $10$ iterations for solving the communication-centric problem. The sum-rate increases from $1.63$ bps/Hz to $68.49$ bps/Hz, which shows the effectiveness and the great improvement in the communication performance of our proposed algorithm.
\subsection{Performance Comparison for Different Setups}
\begin{figure}[!t]
\centering
\begin{subfigure}[b]{\linewidth}
    \includegraphics[width=\linewidth]{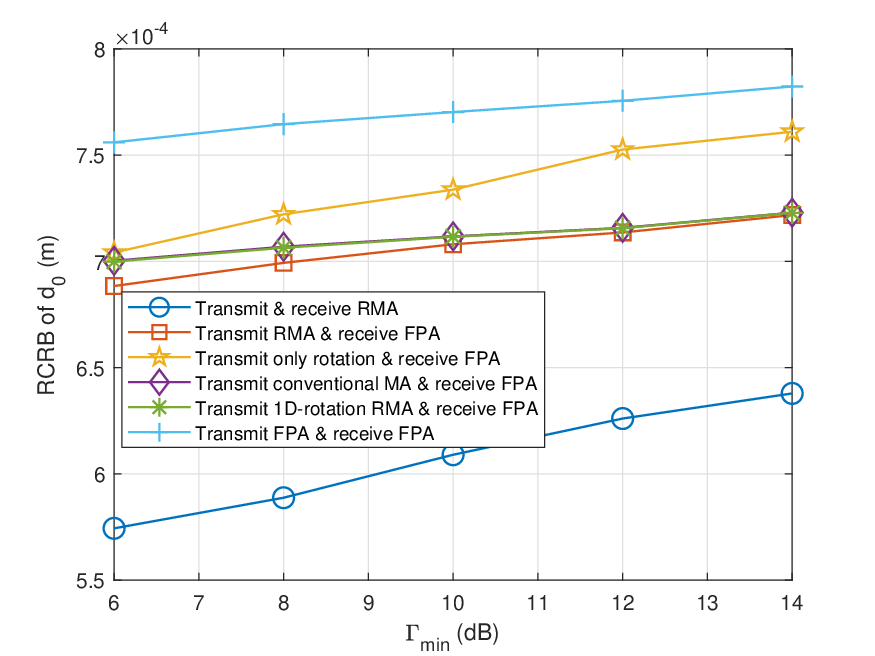}
    \caption{RCRB of $d_0$.}
    \label{modesd}
\end{subfigure}
\begin{subfigure}[b]{\linewidth}
    \includegraphics[width=\linewidth]{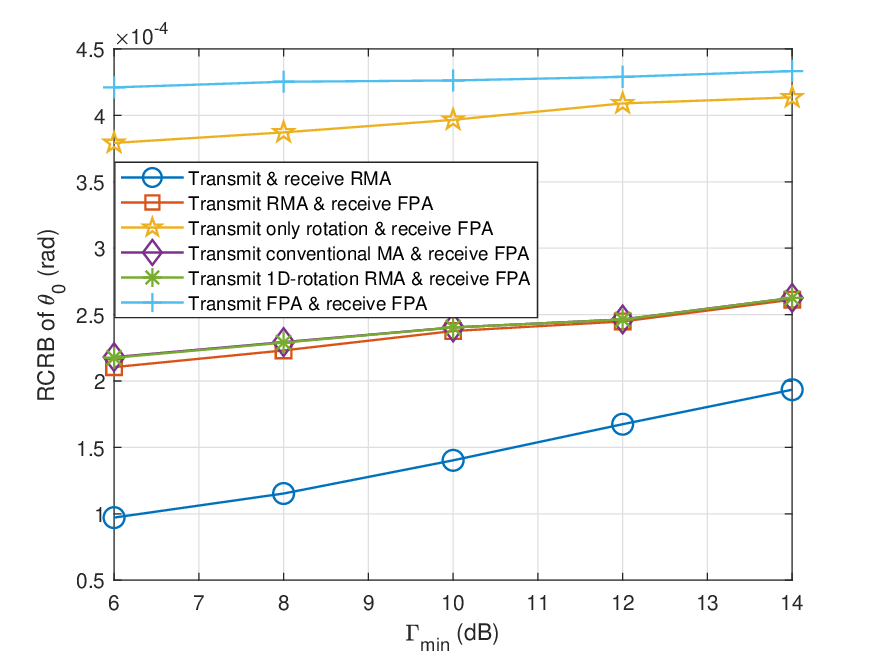}
    \caption{RCRB of $\theta_0$.}
    \label{modestheta}
\end{subfigure}
\begin{subfigure}[b]{\linewidth}
    \includegraphics[width=\linewidth]{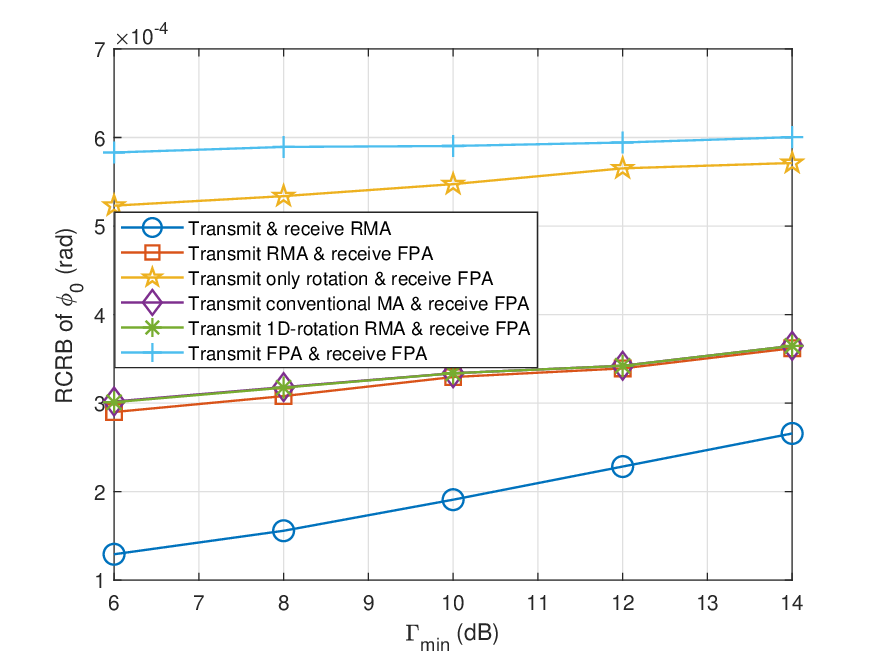}
    \caption{RCRB of $\phi_0$.}
    \label{modesphi}
\end{subfigure}
\caption{\justifying RCRB versus $\Gamma_{\min}$.}
\label{modesrcrb}
\end{figure}
\begin{figure}[!t]
  \centering
    \includegraphics[width=\linewidth]{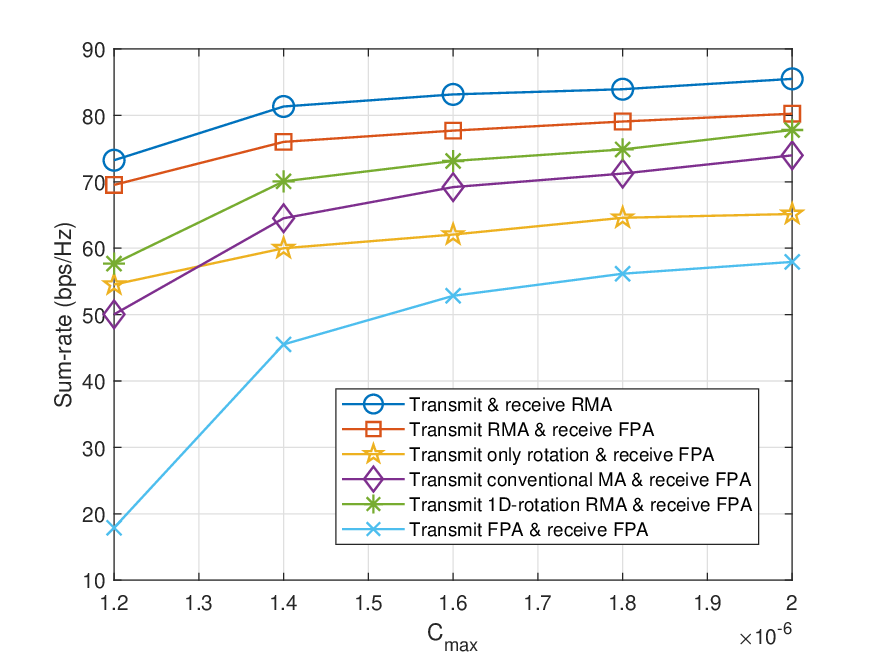}
    \caption{\justifying Sum-rate versus $C_{\max}$.}
    \label{modessr}
\end{figure}
\begin{figure}[!t]
\centering
\begin{subfigure}[b]{\linewidth}
    \includegraphics[width=\linewidth]{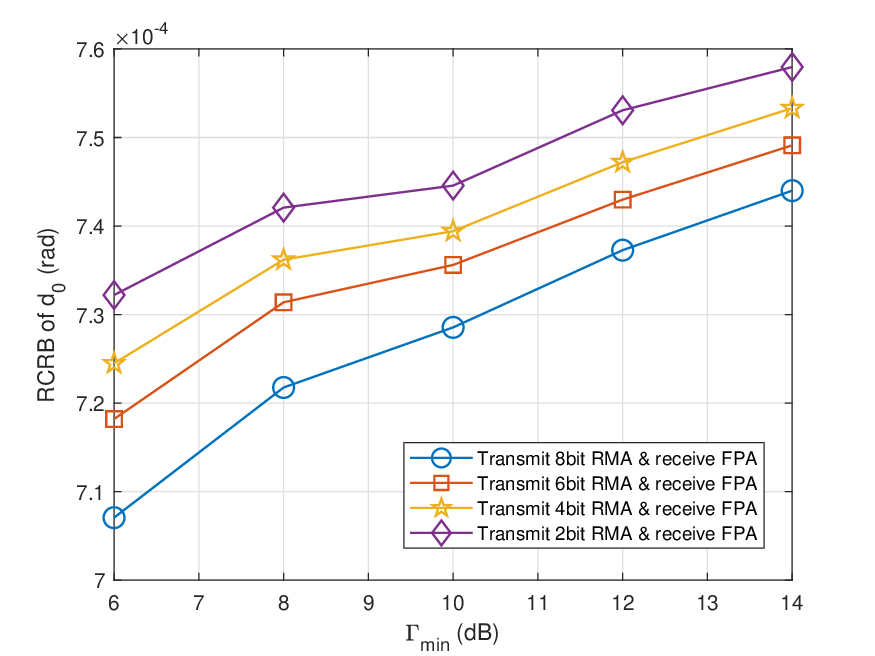}
    \caption{RCRB of $d_0$.}
    \label{bitsd}
\end{subfigure}
\begin{subfigure}[b]{\linewidth}
    \includegraphics[width=\linewidth]{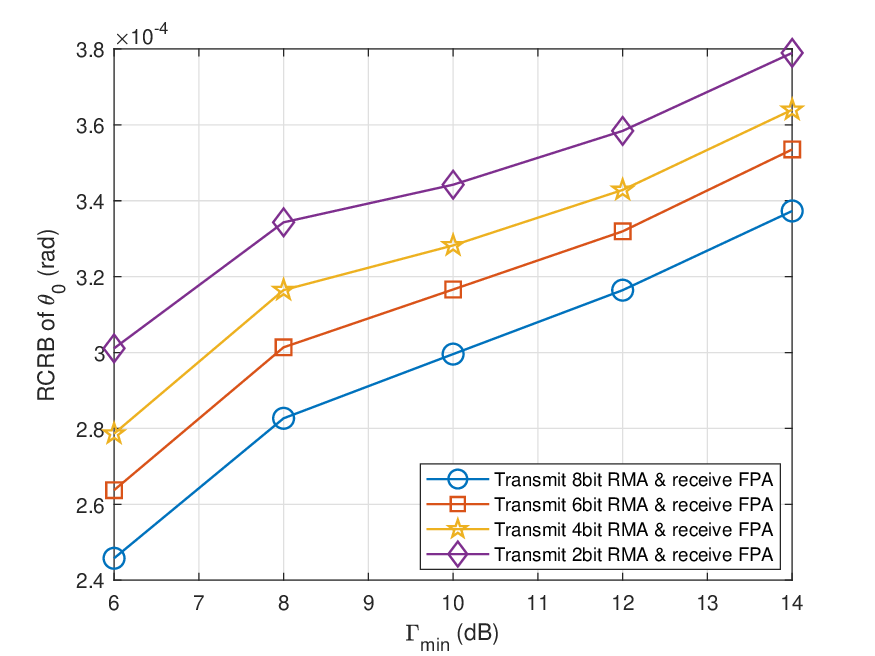}
    \caption{RCRB of $\theta_0$.}
    \label{bitstheta}
\end{subfigure}
\begin{subfigure}[b]{\linewidth}
    \includegraphics[width=\linewidth]{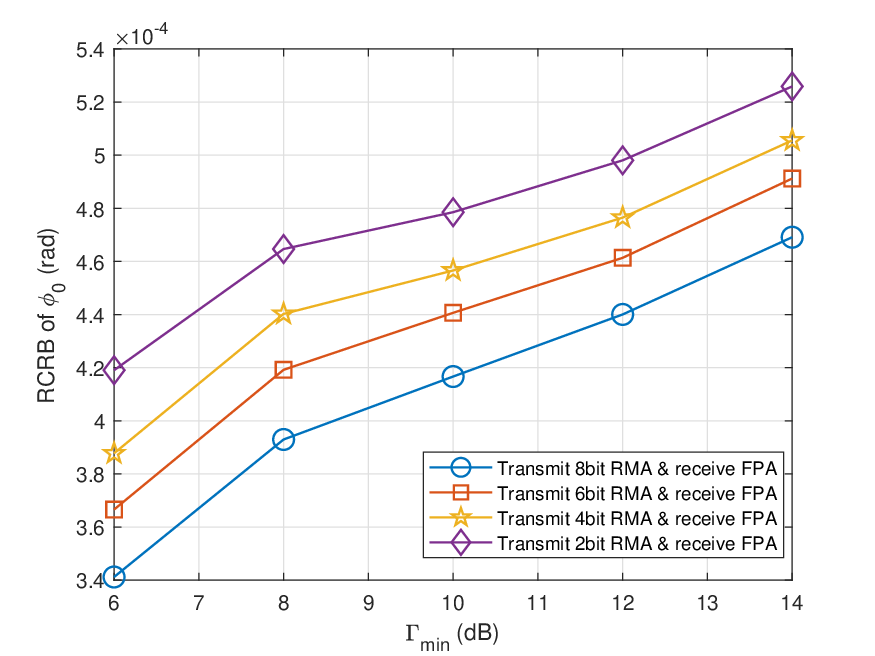}
    \caption{RCRB of $\phi_0$.}
    \label{bitsphi}
\end{subfigure}
\caption{\justifying RCRB versus $\Gamma_{\min}$ with discrete rotations.}
\label{bitsrcrb}
\end{figure}
\begin{figure}[!t]
  \centering
    \includegraphics[width=\linewidth]{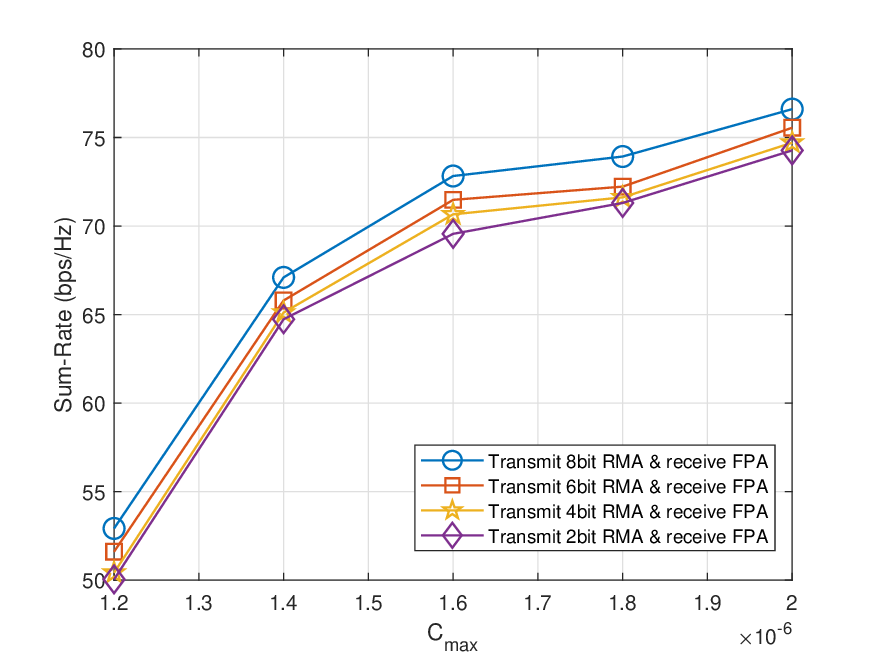}
    \caption{Sum-rate versus $C_{\max}$ with discrete rotations.}
    \label{bitssr}
\end{figure}
To demonstrate the individual effect and synergy of RMAs' rotations and element position movement, we introduce the following six different setups for the antennas at BS:
\begin{itemize}
  \item Transmit \& receive RMA:~Both the TP and RP can execute 3D rotations while all the transmit and receive antennas can move in the 2D planes.
  \item Transmit RMA \& receive FPA:~The TP can execute 3D rotations while the transmit antennas can move in TP. The rotation angles of RP and the positions of receive antennas are fixed.
  \item Transmit only rotation \& receive FPA:~The TP can execute 3D rotations. The rotation angles of RP, the positions of transmit and receive antennas are fixed.
  \item Transmit conventional MA \& receive FPA:~The transmit antennas can move in TP. The rotation angles of TP and RP, the positions of receive antennas are fixed.
  \item Transmit 1D-rotation RMA \& receive FPA:~The TP can execute 1D rotations, i.e., $\alpha^{\text t}$, while the transmit antennas can move in TP. The rotation angles of RP and the positions of receive antennas are fixed.
  \item Transmit FPA \& receive FPA:~The rotation angles of TP and RP as well as the positions of transmit and receive antennas are fixed.
\end{itemize}

Fig. \ref{modesrcrb} shows the sensing performance comparison for different $\Gamma_{\min}$. As can be observed, the RCRBs of $d_0$, $\theta_0$, and $\phi_0$ increase as increment of $\Gamma_{\min}$. When $\Gamma_{\min}$ increases, higher power is required in the beams serving communication users due to stricter SINR constraints and the RMAs also move and rotate accordingly to the positions and angles that are more inclined towards ensuring communication performance. Compared to other setups, the setup with transmit \& receive RMA significantly achieves the best sensing performance, which indicates a synergy between transmit and receive RMAs. The setups with transmit RMA \& receive FPA, transmit conventional MA \& receive FPA, and transmit 1D-rotation RMA \& receive FPA achieve comparable sensing performance. It implies that the sensing performance gain from rotations is relatively limited in this context. Comparing the performance of the setups with transmit only rotation \& receive FPA and transmit FPA \& receive FPA, we can find that the rotations of RMA perform a higher gain for RCRB of $d_0$, since the rotations allow for a more noticeable difference in the distance between the antennas at BS and the sensing target.

Fig. \ref{modessr} shows the communication performance comparison for different $C_{\max}$ with $\Gamma_{\min}=6$dB. As shown, the downlink sum-rate increases as the increase of $C_{\max}$ and gradually slows down, which can be regarded as a manifestation of the CRB-rate Pareto boundary \cite{Xiong2022}. In Fig. \ref{modessr}, the performance gain from rotations is more significant compared to the sensing-centric design because the effective aperture loss is mitigated by the rotations of TP and RP. The downlink sum-rate of the setup with transmit \& receive RMA is substantially higher than transmit FPA \& receive FPA. At $C_{\max}=1.4\times10^{-6}$, the former achieves $78.7$\% performance improvement over the latter. This indicates that the proposed BS design and algorithm are desirable and appealing when the ISAC network is heavily loaded.

\subsection{Performance for Discrete Rotations}
In practical deployment, continuous rotation of RMAs is difficult to realize since RMAs need to be mechanically moved by physical devices, such as a stepper motor that can only adjust the rotations of each RMA in discrete steps, which are usually specified by the motor used \cite{Shao20242}. As such, we employ discrete rotations for the transmit RMA \& receive FPA setup in Fig. \ref{bitsrcrb} and Fig. \ref{bitssr}. {\color{black}Specifically, we uniformly discretize the rotation angles with $2$ to $8$ bits by quantizing the corresponding parts of each particle obtained by \eqref{xis} in Algorithm \ref{Algorithm1} and Algorithm \ref{Algorithm2}.} From Fig. \ref{bitsd}-\ref{bitsphi}, it is obvious that quantization reduces the sensing performance. As the number of quantization bits decreases, the system performance exhibits a gradual degradation. With increasing $\Gamma_{\min}$, the sensing performance for target's position information consistently declines, indicated by sustained increase of RCRB. This characterizes the Pareto boundary of sensing and communication performance under discrete transmit antenna rotation angle setup. As shown in Fig. \ref{bitssr}, quantization has a more significant impact on the reduction of communication performance. As the number of quantization bits decreases, the downlink sum-rate also gradually decreases, even falling below the setup with transmit 1D continuous rotation RMA \& receive FPA. Similarly, as $C_{\max}$ increases, the growth of the downlink sum-rate gradually slows down, which reflects the CRB-rate Pareto boundary under discrete transmit antenna rotation angle setup. Fig. \ref{bitsrcrb} and Fig. \ref{bitssr} indicate that sufficient quantization level of RMA discrete rotation should be chosen based on the performance requirement and hardware conditions in both sensing-centric and communication-centric systems.

\section{Conclusion}\label{s6}
In this paper, we investigated the RMA-enabled near-field ISAC system. We derived an accurate near-field channel model by introducing the effective aperture loss to the conventional spherical wave model and formulated two optimization problems for the sensing-centric and communication-centric designs, respectively. To solve the non-convex sensing-centric problem, we proposed an AO-based algorithm with the SDR method for optimizing the transmit beamforming and the PSO method for optimizing RMAs' rotations and positions. Moving on to the communication-centric problem, we first rewrote it into a more tractable form and then proposed a similar algorithm to optimize the transmit beamforming and RMAs' rotations and positions. Numerical results verified the effectiveness of the proposed algorithms and the advantages of the considered RMA-enabled ISAC network compared to conventional FPA-based systems. {\color{black}The proposed system is expected to achieve more accurate sensing and higher downlink communication rate across a variety of scenarios. In particular, we foresee that the integration of RMA will open up new opportunities in massive MIMO, unmanned aerial vehicle communication, and adaptive multiple access. These directions deserve to be further explored in the future.}
\begin{appendix}
\subsection{Fisher Information Matrices for Near-Filed ISAC}\label{App1}
For a given $\mathbf{G}=\eta\mathbf{g}_0\mathbf{h}_0^{\mathsf T}$, the unknown parameters are denoted by $\bm{\zeta}=[d_0,\theta_0,\phi_0,\eta^{\text r},\eta^{\text i}]$. The covariance matrix of the received ISAC signal is represented by $\mathbf{R}_x$. Define $\mathbf{Y}_0=[\mathbf{y}_0(1),\dots,\mathbf{y}_0(T)]\in\mathbb{C}^{N_{\text r}\times T}$ and $\mathbf{u}_0={\text{vec}}(\mathbf{Y}_0)$. Then, according to \cite{Kay1993}, the FIM for estimating $\bm{\zeta}$ can be partitioned as
\begin{align}
\mathbf{J}_{\bm{\zeta}}=\frac{2}{\sigma_0^2}\Re\left\{\frac{\partial \mathbf{u}_0^{\mathsf H}}{\partial\bm{\zeta}}\frac{\partial \mathbf{u}_0}{\partial\bm{\zeta}}\right\}=\begin{bmatrix}
\mathbf{J}_{11}&\mathbf{J}_{12}\\
\mathbf{J}_{12}^{\mathsf T}&\mathbf{J}_{22}
\end{bmatrix},
\end{align}where $\mathbf{J}_{11}=\begin{bmatrix}
\mathbf{J}_{d_0d_0}&\mathbf{J}_{d_0\theta_0}&\mathbf{J}_{d_0\phi_0}\\
\mathbf{J}_{\theta_0d_0}&\mathbf{J}_{\theta_0\theta_0}&\mathbf{J}_{\theta_0\phi_0}\\
\mathbf{J}_{\phi_0d_0}&\mathbf{J}_{\phi_0\theta_0}&\mathbf{J}_{\phi_0\phi_0}
\end{bmatrix}$,
$\mathbf{J}_{12}=\begin{bmatrix}
\mathbf{J}_{d_0\eta^{\text r}}&\mathbf{J}_{d_0\eta^{\text i}}\\
\mathbf{J}_{\theta_0\eta^{\text r}}&\mathbf{J}_{\theta_0\eta^{\text i}}\\
\mathbf{J}_{\phi_0\eta^{\text r}}&\mathbf{J}_{\phi_0\eta^{\text i}}
\end{bmatrix}$, and
$\mathbf{J}_{22}=\begin{bmatrix}
\mathbf{J}_{\eta^{\text r}\eta^{\text r}}&0\\
0&\mathbf{J}_{\eta^{\text i}\eta^{\text i}}
\end{bmatrix}$. For $a,b\in\{d_0,\theta_0,\phi_0,\eta^{\text r},\eta^{\text i}\}$, the value of each entry is given by
\begin{align}
\mathbf{J}_{ab}=\frac{2}{\sigma_0^2}\Re\left\{\frac{\partial \mathbf{u}_0^{\mathsf H}}{\partial a}\frac{\partial \mathbf{u}_0}{\partial b}\right\}.
\end{align}
By exploiting the approximation $\mathbf{R}_x\approx\frac{1}{T}\mathbf{X}\mathbf{X}^{\mathsf H}$, we have
\begin{align}
&\mathbf{J}_{pq}=\frac{2|\eta|^2T}{\sigma_0^2}\Re\left\{{\text{tr}}(\frac{\partial\mathbf{g}_0\mathbf{h}_0^{\mathsf T}}{\partial q}\mathbf{R}_x\frac{\partial\mathbf{h}_0^*\mathbf{g}_0^{\mathsf H}}{\partial p})\right\},\\
&\mathbf{J}_{p\eta^{\text r}}=\frac{2T}{\sigma_0^2}\Re\left\{\eta^*{\text{tr}}({\mathbf{g}_0\mathbf{h}_0^{\mathsf T}}\mathbf{R}_x\frac{\partial\mathbf{h}_0^*\mathbf{g}_0^{\mathsf H}}{\partial p})\right\},\\
&\mathbf{J}_{p\eta^{\text i}}=\frac{2T}{\sigma_0^2}\Re\left\{{\text j}\eta^*{\text{tr}}({\mathbf{g}_0\mathbf{h}_0^{\mathsf T}}\mathbf{R}_x\frac{\partial\mathbf{h}_0^*\mathbf{g}_0^{\mathsf H}}{\partial p})\right\},\\
&\mathbf{J}_{\eta^{\text r}\eta^{\text r}}=\mathbf{J}_{\eta^{\text i}\eta^{\text i}}=\frac{2T}{\sigma_0^2}\Re\left\{{\text{tr}}({\mathbf{g}_0\mathbf{h}_0^{\mathsf T}}\mathbf{R}_x{\mathbf{h}_0^*\mathbf{g}_0^{\mathsf H}})\right\},
\end{align}for $p,q\in\{d_0,\theta_0,\phi_0\}$.

\subsection{Proof of Proposition \ref{theorem0}}\label{App2}
Let $\tilde{\mathbf{R}}_x$, $\tilde{\bm{\Omega}}_1,\dots,\tilde{\bm{\Omega}}_K$ be an arbitrary global optimum of $\mathcal{P}_3$. Our objective is to construct another global optimum, denoted as $\mathbf{R}_x^\circ$, ${\bm{\Omega}}_1^\circ,\dots,{\bm{\Omega}}_K^\circ$, such that $\mathbf{R}_x^\circ$ is positive semidefinite, and ${\bm{\Omega}}_k^\circ$ is positive semidefinite and rank-one, for $k=1,\dots,K$. To achieve this, we prove the theorem by constructing $\mathbf{R}_x^\circ$, ${\bm{\Omega}}_1^\circ,\dots,{\bm{\Omega}}_K^\circ$ from $\tilde{\mathbf{R}}_x$, $\tilde{\bm{\Omega}}_1,\dots,\tilde{\bm{\Omega}}_K$ with $\mathbf{R}_x^\circ=\tilde{\mathbf{R}}_x$, $\mathbf{w}_k^\circ=\left(\mathbf{h}_k^{\mathsf T}\tilde{\bm{\Omega}}_k\mathbf{h}_k^{*}\right)^{-\frac{1}{2}}\tilde{\bm{\Omega}}_k\mathbf{h}_k^{*}$, and ${\bm{\Omega}}_k^\circ=\mathbf{w}_k^\circ(\mathbf{w}_k^\circ)^{\mathsf H}$, for $k=1,\dots,K$.

{\color{black}We now prove that $\mathbf{R}_x^\circ$, ${\bm{\Omega}}_1^\circ,\dots,{\bm{\Omega}}_K^\circ$ is also a global optimum to $\mathcal{P}_3$. Since the objective function is not directly determined by $\mathbf{R}_x^\circ$, ${\bm{\Omega}}_1^\circ,\dots,{\bm{\Omega}}_K^\circ$ while \eqref{pc} and \eqref{Uc} hold for $\mathbf{R}_x^\circ=\tilde{\mathbf{R}}_x$, we only need to validate that $\mathbf{R}_x^\circ$, ${\bm{\Omega}}_1^\circ,\dots,{\bm{\Omega}}_K^\circ$ is also a feasible solution to $\mathcal{P}_3$ by proving \eqref{Rc2} and \eqref{SINRc2} hold.

First, for any vector $\bm{\vartheta}\in\mathbb{C}^{N_{\text{t}}\times 1}$, we have
\begin{align}
\bm{\vartheta}^{\mathsf H}(\tilde{\bm{\Omega}}_k-{\bm{\Omega}}_k^\circ)\bm{\vartheta}=\bm{\vartheta}^{\mathsf H}\tilde{\bm{\Omega}}_k\bm{\vartheta}-\left(\mathbf{h}_k^{\mathsf T}\tilde{\bm{\Omega}}_k\mathbf{h}_k^{*}\right)^{-1}|\bm{\vartheta}^{\mathsf H}\tilde{\bm{\Omega}}_k\mathbf{h}_k^{*}|^2.
\end{align}
According to the Cauchy-Schwarz inequality, we can derive
\begin{align}
\left(\mathbf{h}_k^{\mathsf T}\tilde{\bm{\Omega}}_k\mathbf{h}_k^{*}\right)\left(\bm{\vartheta}^{\mathsf H}\tilde{\bm{\Omega}}_k\bm{\vartheta}\right)\geq |\bm{\vartheta}^{\mathsf H}\tilde{\bm{\Omega}}_k\mathbf{h}_k^{*}|^2,
\end{align}so $\bm{\vartheta}^{\mathsf H}(\tilde{\bm{\Omega}}_k-{\bm{\Omega}}_k^\circ)\bm{\vartheta}\geq 0$ holds for any vector $\bm{\vartheta}\in\mathbb{C}^{N_{\text{t}}\times 1}$. Therefore,
$\mathbf{R}_x^\circ=\tilde{\mathbf{R}}_x\succeq\sum_{k=1}^{K}\tilde{\bm{\Omega}}_k\succeq\sum_{k=1}^{K}{\bm{\Omega}}_k^\circ$,
namely constraint \eqref{Rc2} holds for $\mathbf{R}_x^\circ$, ${\bm{\Omega}}_1^\circ,\dots,{\bm{\Omega}}_K^\circ$.} Moreover, we can derive that
\begin{align}
\mathbf{h}_k^{\mathsf T}\bm{\Omega}_k^\circ\mathbf{h}_k^*=\mathbf{h}_k^{\mathsf T}\mathbf{w}_k^\circ({\mathbf{w}_k^\circ})^{\mathsf H}\mathbf{h}_k^*=\mathbf{h}_k^{\mathsf T}\tilde{\bm{\Omega}}_k\mathbf{h}_k^*.
\end{align}Thus
\begin{align}
&\frac{1+\Gamma_{\min}}{\Gamma_{\min}}\mathbf{h}_k^{\mathsf T}\bm{\Omega}_k^\circ\mathbf{h}_k^{*}=\frac{1+\Gamma_{\min}}{\Gamma_{\min}}\mathbf{h}_k^{\mathsf T}\tilde{\bm{\Omega}}_k\mathbf{h}_k^{*}\nonumber\\
&\geq\mathbf{h}_k^{\mathsf T}\tilde{\mathbf{R}}_x\mathbf{h}_k^{*}+\sigma_k^2=\mathbf{h}_k^{\mathsf T}\mathbf{R}_x^\circ\mathbf{h}_k^{*}+\sigma_k^2,
\end{align}which ensures that constraint \eqref{SINRc2} holds for $\mathbf{R}_x^\circ$, ${\bm{\Omega}}_1^\circ,\dots,{\bm{\Omega}}_K^\circ$.

 With the derivation above, it is verified that $\mathbf{R}_x^\circ$, ${\bm{\Omega}}_1^\circ,\dots,{\bm{\Omega}}_K^\circ$ is also a feasible solution, and furthermore, a global optimum to $\mathcal{P}_3$, which completes the proof.
% $\frac{\mathbf{h}_k^{\mathsf T}\mathbf{w}_k^\circ({\mathbf{w}_k^\circ})^{\mathsf H}\mathbf{h}_k^*}{\mathbf{h}_k^{\mathsf T}(\mathbf{R}_x^\circ-\bm{\Omega}_k^\circ)\mathbf{h}_k^{\mathsf *}+\sigma_k^2}=\frac{\mathbf{h}_k^{\mathsf T}\tilde{\bm{\Omega}}_k\mathbf{h}_k^*}{\mathbf{h}_k^{\mathsf T}(\tilde{\mathbf{R}}_x-\tilde{\bm{\Omega}}_k)\mathbf{h}_k^{\mathsf *}+\sigma_k^2}$

\end{appendix}

\end{document}